\documentclass[reprint,aps,prl,amsmath,amssymb,longbibliography, superscriptaddress]{revtex4-2}
\usepackage{graphicx}
\usepackage{physics}
\usepackage{multirow}
\usepackage{tabularx,booktabs,ragged2e,array}     
\usepackage{booktabs}
\usepackage{ragged2e}   
\usepackage[table]{xcolor}
\usepackage{todonotes}
\usepackage[colorlinks, linkcolor= blue, citecolor = blue, urlcolor=blue]{hyperref}
\usepackage{pifont}
\usepackage{cancel}
\usepackage{comment}
\usepackage{mathrsfs}
\usepackage{bm}
\usepackage[normalem]{ulem}
\usepackage{subcaption}
\usepackage{nameref}

\def\bea{\begin{eqnarray}}
\def\eea{\end{eqnarray}}
\def\be{\begin{equation}}
\def\ee{\end{equation}}
\def\tc{\textcolor}

\def\kb{{\bm k}}

\definecolor{mygreen}{HTML}{E7FF86}
\begin{document}

\title{Quantum oscillation spectroscopy of 
Fermi-surface topologies in tetralayer graphene}


\author{Abhijit Halder}
\affiliation{Department of Physics, Indian Institute of Science, Bangalore, 560012, India}

\author{Harsh Varshney}
\affiliation{Department of Physics, Indian Institute of Technology, Kanpur-208016, India.}

\author{Snehamoyee Hazra}
\affiliation{Department of Physics, Indian Institute of Science, Bangalore, 560012, India}

\author{Santu Kumar Bera}
\affiliation{Department of Physics, Indian Institute of Science, Bangalore, 560012, India}

\author{Souvik Chakraborty}
\affiliation{Department of Physics, Indian Institute of Science, Bangalore, 560012, India}

\author{Ujjal Roy}
\affiliation{Department of Physics, Indian Institute of Science, Bangalore, 560012, India}

\author{Takashi Taniguchi}
\affiliation{Research Center for Materials Nanoarchitectonics, National Institute for Materials Science,
1-1 Namiki, Tsukuba 305-0044, Japan}

\author{Kenji Watanabe}
\affiliation{Research Center for Electronic and Optical Materials, National Institute for Materials
Science, 1-1 Namiki, Tsukuba 305-0044, Japan}

\author{Amit Agarwal}
\email{amitag@iitk.ac.in}
\affiliation{Department of Physics, Indian Institute of Technology, Kanpur-208016, India.}

\author{Anindya Das}
\email{anindya@iisc.ac.in}
\affiliation{Department of Physics, Indian Institute of Science, Bangalore, 560012, India}

\begin{abstract} 
Quantum oscillations offer a direct probe of Fermi-surface topology and electronic degeneracy, yet disentangling both simultaneously across the Lifshitz transitions of multiband systems has remained an open experimental challenge. Here, we use Shubnikov-de Haas spectroscopy on a high-mobility, dual-gated Bernal-stacked tetralayer graphene (B-4LG) device to quantitatively reconstruct the complete sequence of 
six distinct Fermi-surface topologies-gully, annular, singly connected, and multiband pockets. 
The extracted oscillation frequencies determine the extremal momentum-space areas and their spin, valley, and gully-resolved degeneracies, in quantitative agreement with our tight-binding calculations. We further show that a perpendicular magnetic-field, combined with displacement-field lifts the valley degeneracy through an orbital-Zeeman coupling, producing a single-particle valley splitting of several $meV$, far larger than in bilayer or trilayer graphene. 
Our work demonstrates a framework for tracking Fermi-surface topologies and their flavor degeneracies in multiband quantum materials.
\end{abstract}

\maketitle

\tc{blue}{\textit{Introduction:---}}
Fermi-surface topology governs the low-energy properties of metals and semimetals. Changes in Fermi surface topology, known as Lifshitz transitions, can dramatically modify electronic responses without involving spontaneous symmetry breaking~\cite{lifshitz1960anomalies,Volovik_2018}. Quantum oscillations provide a precise probe of extremal Fermi-surface areas. However, quantitatively reconstructing both the topology of the underlying Fermi pockets and their associated spin, valley, or other flavor degeneracies remains a major experimental challenge~\cite{vignolle2008quantum,PhysRevLett.114.147001,liu2010evidence,Ren2017SciAdv,yelland2011high,PhysRevLett.110.256403,PhysRevLett.124.086601,yang2019topological,Liu2020PNAS,shi2017enhanced,VARLET201519,PhysRevLett.120.096802,jayaraman2021evidence,Datta2024nonlinear,Ahmed2025,Sahani2025OMR}.

Graphene multilayers provide an ideal platform for addressing this challenge because their low-energy electronic structure can be continuously tuned by carrier density and displacement field~\cite{castro2009electronic, mccann2013electronic,PhysRevB.79.035421}. Interlayer hopping, trigonal warping, and electrostatic asymmetry generate a rich sequence of Lifshitz transitions~\cite{VARLET201519,PhysRevLett.120.096802,doi:10.1021/acs.nanolett.4c01133,Harsh2026asymmetric,PhysRevB.101.245411,PhysRevLett.121.167601,nm8b-5vgm}. Bernal-stacked tetralayer graphene (B-4LG) is particularly attractive because its low-energy spectrum contains coexisting light- and heavy-mass bilayer-like bands. Their interplay with trigonal warping and displacement-field-induced hybridization produces an unusually rich hierarchy of gully, annular, singly connected, and multiband Fermi surfaces within experimentally accessible energies. Previous transport, compressibility, and magnetic-focusing measurements have revealed signatures of a few Lifshitz transitions and band anisotropy~\cite{PhysRevLett.120.096802,Qu_2025,Qu_2025_2,PhysRevLett.125.036803,doi:10.1021/acs.nanolett.4c01133, Klanurak20224lg}. 
However, no experiment has quantitatively reconstructed the complete sequence of Fermi-surface topologies together with their associated flavor degeneracies and field-induced valley splitting.

\begin{figure}[t!]
    \centering
    \includegraphics[width=\linewidth]{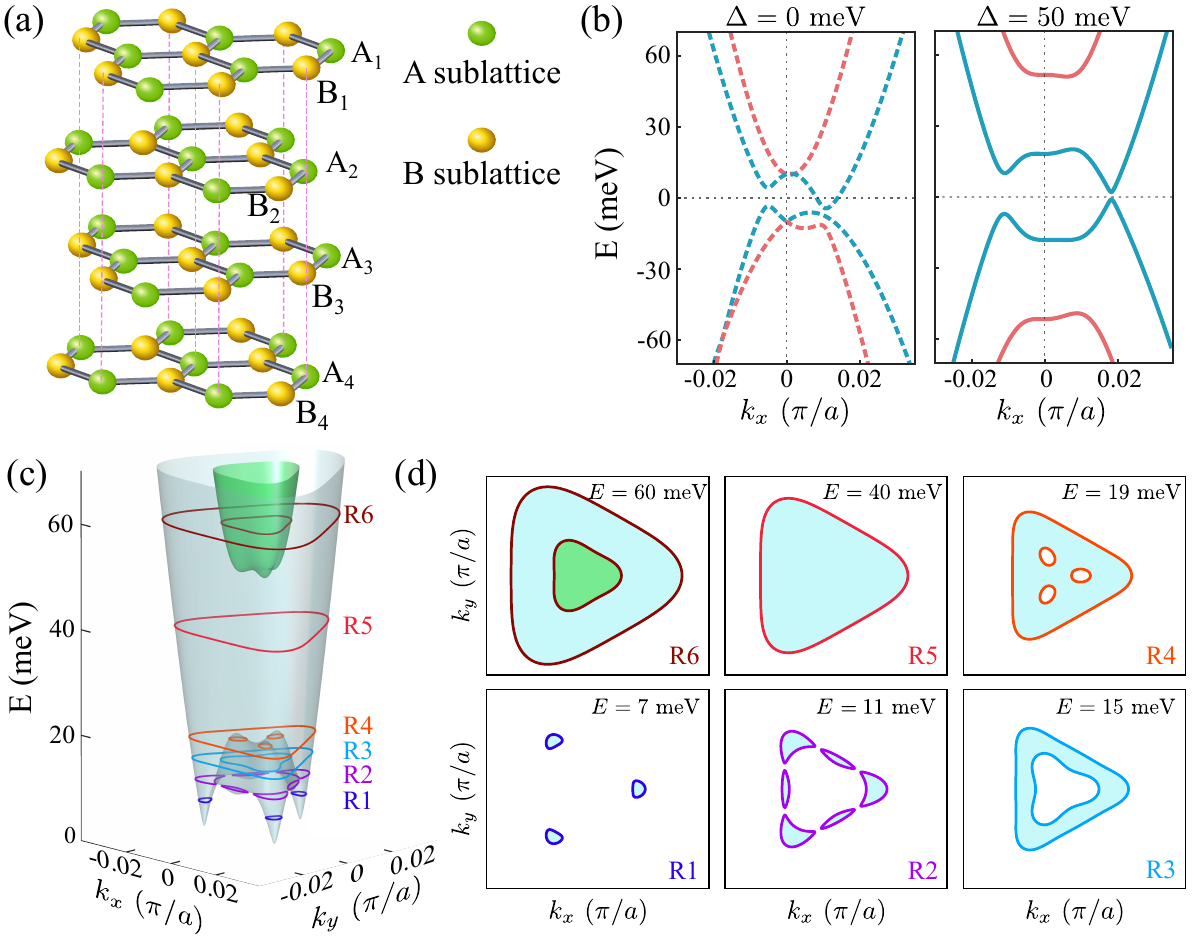}
    \caption{\justifying
    \textbf{Evolution of the Fermi-surface topology in Bernal-stacked tetralayer graphene.}
    (a) Crystal structure of Bernal-stacked (ABAB) tetralayer graphene (B-4LG).
    (b) Low-energy band structure near the $K$ point at zero and finite displacement fields ($\Delta$). A finite displacement field breaks inversion symmetry, hybridizes the light- and heavy-mass bilayer-like bands, and opens gaps at their crossings.
    (c) Three-dimensional conduction-band dispersion illustrating the strong anisotropy induced by trigonal warping. (d) Constant-energy contours showing six conduction-band Fermi-surface topologies separated by Lifshitz transitions. The interplay of trigonal warping, interlayer coupling, and displacement-field-induced hybridization produces gully, annular, singly connected, and multiband Fermi surfaces.
\label{fig1}}
\end{figure}

Here, we use Shubnikov-de Haas (SdH) oscillation spectroscopy~\cite{Onsager01091952,lifshitz1956theory,sahani2026quantum}, for the first time to quantitatively reconstruct both the Fermi-surface topology and flavor degeneracy of a high-mobility dual-gated B-4LG device across a sequence of Lifshitz transitions. We identify six distinct 
Fermi-surface topologies and determine their extremal momentum-space areas together with their spin, valley, and gully degeneracies from the measured SdH frequencies. We further show that the interplay of finite displacement and magnetic field lifts the valley degeneracy. This produces a pronounced splitting of the single-particle Landau-level spectrum and a corresponding evolution of the incompressible quantum Hall states. The measured oscillation frequencies, degeneracy reconstruction, and valley splitting are all in excellent agreement with our tight-binding calculations, providing a quantitative Fermi-surface and flavor-degeneracy map of Bernal-stacked tetralayer graphene.

\tc{blue}{\textit{Series of Lifshitz transitions:—}}
As shown in Fig.~\ref{fig1}(a), B-4LG consists of two Bernal bilayer graphene (BLG) subsystems coupled by inter-bilayer hopping. Its low-energy spectrum contains four bands near charge neutrality [Fig.~\ref{fig1}(b)]~\cite{PhysRevLett.125.036803, doi:10.1021/acs.nanolett.4c01133, chen2023gate}. The corresponding tight-binding model is described in End Matter~\hyperref[4lg_tb]{EM1}. Interlayer skew hopping produces strong trigonal warping, giving rise to anisotropic dispersions with multiple extrema and saddle points. A finite displacement field breaks inversion symmetry and hybridizes the light- and heavy-mass bilayer-like bands. This leads to a reconstruction of the low-energy Fermi surface into multiple pockets, including Dirac gullies within each valley~\cite{PhysRevB.80.165409, avetisyan2009electric, PhysRevB.106.L161405}. 

As the Fermi energy is varied, the Fermi surface undergoes a cascade of Lifshitz transitions~\cite{PhysRevLett.120.096802, doi:10.1021/acs.nanolett.4c01133, ren2025electric}. Figures~\ref{fig1}(c,d) show a sequence of six distinct conduction-band Fermi-surface topologies. Starting from a multiband Fermi surface at higher energies (R6), the contours evolve into a singly connected pocket (R5), followed by annular Fermi surfaces (R4 and R3), and finally disconnected pockets (R2 and R1). These successive topological reconstructions result from the combined effects of interlayer coupling, displacement-field induced band hybridization, and trigonal warping.

\begin{figure}[t]
\centering
\includegraphics[width=\linewidth]{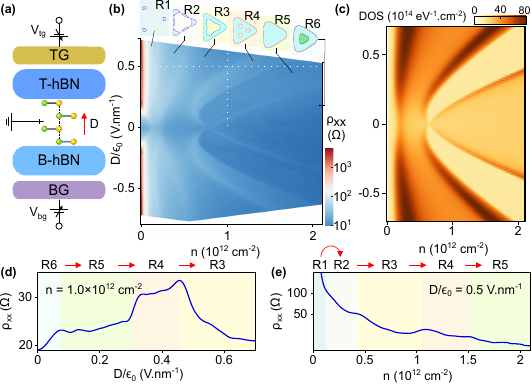}
\caption{\justifying \textbf{Transport signatures of Lifshitz transitions in B-4LG.}
{(a) Schematic of a dual-gated, hBN-encapsulated B-4LG Hall-bar device with independent control of carrier density ($n$) and displacement field ($D$).}
(b) Longitudinal resistivity, $\rho_{xx}$, as a function of $n$ and $D$, measured at $T=20~\mathrm{mK}$ and zero magnetic field. Resistivity anomalies separate the different Fermi-surface topology regimes, shown in the insets. White dashed lines indicate the cuts in (d,e). 
{(c) Calculated density of states (DOS) as a function of $n$ and $D$. The DOS features track the transport anomalies measured in (b).}
(d,e) Representative line cuts of $\rho_{xx}$ as a function of $D/\epsilon_0$ at fixed  $n=1.0\times10^{12}~\mathrm{cm^{-2}}$ (d) and as a function of $n$ at fixed $D/\epsilon_0=0.5~\mathrm{V\,nm^{-1}}$ (e).
} 
\label{fig2}
\end{figure}

\tc{blue}{\textit{Transport signatures of Fermi-surface topology:---}}
To probe these Fermi-surface reconstructions, we fabricate high-mobility, dual-gated B-4LG Hall-bar devices using a dry-transfer technique [Fig.~\ref{fig2}(a)]. The top and back gates independently control the carrier density ($n$) and the vertical displacement field ($D$).

Figure~\ref{fig2}(b) shows the measured longitudinal resistivity, $\rho_{xx}$, as a function of $n$ and $D$ on the electron-doped side at zero magnetic field. The corresponding hole-side data are presented in Sec.~IV of the supplementary material (SM)~\cite{SM}.
Pronounced resistivity anomalies appear throughout the phase diagram and closely follow the features of the tight-binding density of states (DOS) [Fig.~\ref{fig2}(c)]. Together, these identify the six Fermi-surface regimes illustrated by the insets in Fig.~\ref{fig2}(b). In addition, a narrow region near charge neutrality and small displacement field hosts coexisting electron- and hole-like pockets (see Sec.~IV of the SM~\cite{SM}). 

The line cuts in Figs.~\ref{fig2}(d,e) resolve a sequence of $\rho_{xx}$ extrema as either $D$ or $n$ is varied. Their close correspondence with the calculated DOS confirms that these transport anomalies originate from successive Fermi-surface reconstructions.

\begin{figure*}[t]
\centering
\includegraphics[width=\linewidth]{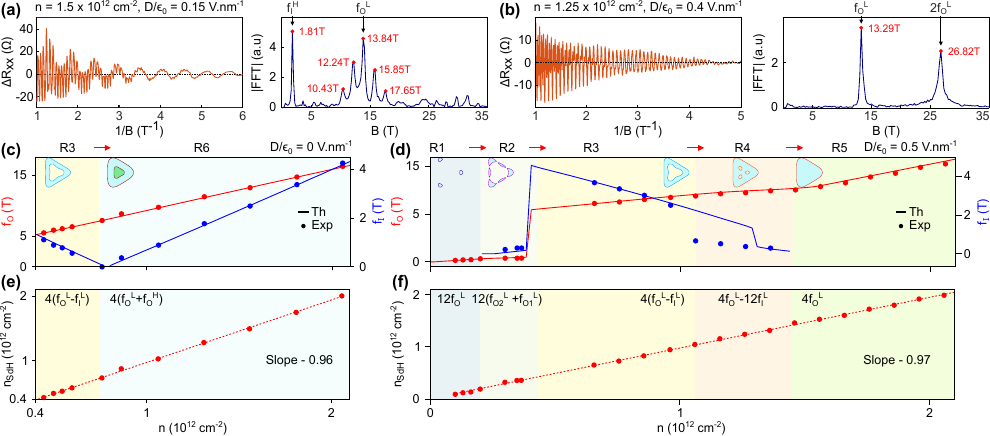}
\caption{\justifying \textbf{SdH mapping of Fermi-surface topology and degeneracy in B-4LG.}
{(a,b) Background-subtracted SdH oscillations ($\Delta R_{xx}$) versus inverse magnetic field ($1/B$) and the corresponding FFT spectra, measured at $n=1.5\times10^{12}~\mathrm{cm^{-2}}$, $D/\epsilon_0=0.15~\mathrm{V\,nm^{-1}}$, and $n=1.25\times10^{12}~\mathrm{cm^{-2}}$, $D/\epsilon_0=0.4~\mathrm{V\,nm^{-1}}$, respectively. Red circles mark the dominant frequencies.}
{(c,d) Extracted outer- and inner-contour frequencies, $f_O$ (red) and $f_I$ (blue), versus carrier density at $D/\epsilon_0=0$ and $0.5~\mathrm{V\,nm^{-1}}$. Circles denote experimentally extracted data, while solid lines denote results from tight-binding calculations. The insets show the corresponding Fermi surfaces.}
{(e,f) Carrier density reconstructed from the SdH frequencies, $n_{\rm SdH}=(e/h)\sum_i g_i f_i$, versus gate-induced density $n$. Insets show the degeneracy-weighted frequency combinations used to calculate $n_{\rm SdH}$. Agreement between $n_{\rm SdH}$ and $n$ establishes the sequence of Fermi-surface topologies and their degeneracies.}}
\label{fig3}
\end{figure*} 

\tc{blue}{\textit{SdH oscillations and Fermi-surface degeneracy:---}}
The resistivity anomalies in Fig.~\ref{fig2} indicate Fermi-surface reconstructions but do not determine the underlying topology and degeneracy of the Fermi surfaces. Quantum oscillations directly probe the extremal Fermi-surface areas, their degeneracies, and the associated Berry phases~\cite{Onsager01091952,lifshitz1956theory,novoselov2005two,Zhang2005,tiwari2022experimental,PhysRevLett.82.2147,PhysRevB.95.035103}. In a perpendicular magnetic field, the electronic spectrum is quantized into Landau levels, giving rise to Shubnikov--de Haas (SdH) oscillations. These are described by the Lifshitz--Kosevich formalism,
\begin{equation}
\Delta \rho_{xx} \propto \sum_i a_i
\cos\left[2\pi\left(\frac{f_i}{B}+\Phi_i\right)\right].
\end{equation}
Here, $a_i$, $f_i$, and $\Phi_i$ denote the oscillation amplitude, frequency, and geometric phase factor associated with the $i$th extremal Fermi-surface contour. The oscillation frequency determines the enclosed momentum-space area through the Onsager relation, $A_i=(2\pi e/\hbar)f_i$, while the corresponding carrier density is $n_i=g_i A_i/(2\pi)^2=(e/h)g_i f_i$, where $g_i$ is the degeneracy of the corresponding Fermi contour. The total carrier density is obtained by summing the contributions from all occupied contours, $n=\sum_i n_i$. For annular Fermi surfaces, the inner contour contributes negatively to the carrier density because it encloses empty states already counted by the outer contour. Thus, SdH oscillations provide a quantitative probe of Fermi-surface topology, including annular Fermi surfaces, and flavor degeneracy.

Figures~\ref{fig3}(a,b) show representative background-subtracted SdH oscillations and the corresponding FFT spectra measured in the multiband (R6) and single-band (R5) regimes, respectively (see Sec.~VI of the SM~\cite{SM}). The FFT spectra contain the fundamental frequencies associated with the extremal Fermi-surface contours, together with weaker harmonic and combination peaks. The dominant frequencies are labeled as $f_{\rm O/I}^{\rm L/H}$, where L (H) denotes the light- (heavy-) mass band and O (I) denotes the outer (inner) Fermi surface contour. In R6 [Fig.~\ref{fig3}(a)], the light- and heavy-band contours produce two dominant frequencies, whereas the reconstructed single-band Fermi surface in R5 [Fig.~\ref{fig3}(b)] produces only one.

Figures~\ref{fig3}(c) and \ref{fig3}(d) show the extracted SdH oscillation frequencies as a function of carrier density at $D/\epsilon_0=0$ and $0.5\,\mathrm{V\,nm^{-1}}$, respectively. At $D=0$, both $f_{\mathrm{I}}$ and $f_{\mathrm{O}}$ agree with the tight-binding calculations over the full density range. At $D/\epsilon_0=0.5\,\mathrm{V\,nm^{-1}}$, the agreement persists across most regimes, except in R4, where the calculated density window is narrower than the measured one. This small deviation at large displacement fields may reflect either limitations of the tight-binding parameterization or the breakdown of well-defined semiclassical cyclotron orbits for the corresponding Fermi surface. In either case, the reconstructed Fermi-surface topology remains unchanged. 

A more stringent test compares the carrier density reconstructed from the SdH frequencies, $n_{\rm SdH}$, with the gate-induced density $n$, as shown in Figs.~\ref{fig3}(e,f). The upper insets show the degeneracy-weighted frequency combinations, $\sum_i (e/h)g_i f_i$, used to obtain $n_{\rm SdH}$. At $D=0$, the SdH measurements probe two distinct Fermi-surface regimes. In the lower-density region, the observed frequencies $(f_O^L,f_I^L)$ originate from the outer and inner contours of an annular heavy-band Fermi surface. Since the inner contour encloses states already counted within the outer contour, its contribution enters with a negative sign in the extraction of $n_{\rm SdH}$. At higher density, the system evolves into a multiband regime where $(f_O^L,f_I^H)$ arise from separate heavy- and light-band contours and therefore contribute additively to $n_{\rm SdH}$. In both regimes, the contours remain fourfold degenerate because of spin and valley symmetry, yielding excellent agreement between $n_{\rm SdH}$ and $n$.

At finite displacement field, $D/\epsilon_0 = 0.5~\mathrm{V\,nm^{-1}}$, the low-density R1 regime consists of three symmetry-equivalent gullies in each valley, giving rise to a 12-fold-degenerate Fermi contour ($f_O^L$), and the extracted $n_{\rm SdH}$ accurately reproduces the gate-induced density. In R2, two distinct 12-fold-degenerate gullies ($f_{O1}^L$ and $f_{O2}^L$) coexist and contribute additively. In R3, the Fermi surface acquires an annular topology similar to that observed at low density for $D=0$. In R4, the reconstructed Fermi surface consists of a fourfold-degenerate outer contour ($f_O^L$) and a 12-fold-degenerate inner annular contour ($f_I^L$), with the inner contour contributing negatively to $n_{\rm SdH}$. Finally, R5 is characterized by a single fourfold-degenerate Fermi contour ($f_O^L$), which again accurately reproduces the gate-induced density. The quantitative agreement between the extracted $n_{\rm SdH}$ and the gate-induced density establishes the complete sequence of Fermi-surface topologies together with their associated flavor degeneracies across the Lifshitz cascade in B-4LG. Additional SdH data at several displacement fields and comparisons with tight-binding calculations are presented in Sec.~VIII of the SM~\cite{SM}.


\begin{figure}[t]
\centering
\includegraphics[width=\linewidth]{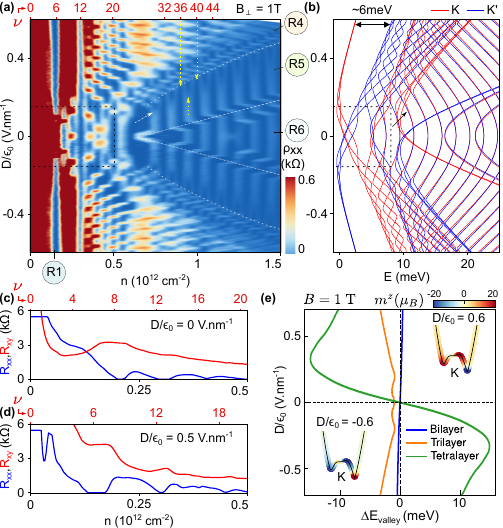}
\caption{\justifying \textbf{Observation of large valley splitting.}
{(a) Measured longitudinal resistivity, $\rho_{xx}$, as a function of carrier density and displacement field at $B_\perp=1~\mathrm{T}$. White and yellow dashed lines trace incompressible Landau-level (LL) sequences in the multiband (R6) and single-pocket (R5) regimes.}
{(b) Calculated single-particle LL spectrum as a function of energy and displacement field. Red and blue lines denote the $K$ and $K'$ valleys, respectively, and all LLs remain spin degenerate. A finite displacement field lifts the valley degeneracy. Black arrows mark the splitting in R5 near the R4 boundary.}
(c,d) Line cuts of $R_{xx}$ (blue) and $R_{xy}$ (red) as a function of carrier density at fixed $D/\epsilon_0=0$ and $0.5~\mathrm{V\,nm^{-1}}$, respectively. The red labels on the upper axes indicate the corresponding filling factors $\nu$. (e) Valley splitting induced by the orbital magnetic moment in B-4LG as a function of displacement field at $B=1~\mathrm{T}$. The splitting is substantially larger than in bilayer and trilayer graphene. Insets show the orbital magnetic moment near the conduction-band edge in the $K$ valley for $D/\epsilon_0=\pm0.6~\mathrm{V\,nm^{-1}}$.}
\label{fig4}
\end{figure} 

\tc{blue}{\textit{Symmetry breaking and valley degeneracy lifting:—}}
Above $B_\perp\approx0.5~\mathrm{T}$, the spectrum in Fig.~\ref{fig3}(b) develops a pronounced second-harmonic peak at $2f_O^L$. This frequency doubling is consistent with lifting the degeneracy of the underlying Fermi-surface contour. Since the Zeeman energy at $B_\perp\approx0.5~\mathrm{T}$ is only $\sim0.05~\mathrm{meV}$, the observed splitting is unlikely to be of spin origin. Instead, it points to valley degeneracy lifting~\cite{PhysRevB.81.115315,PhysRevLett.117.066601}.

To investigate this further, we measure $R_{xx}$ as a function of $n$ and $D$ at several magnetic fields. Figure~\ref{fig4}(a) shows a color map at $B_\perp=1~\mathrm{T}$, where the vertical strips of low $R_{xx}$ correspond to incompressible Landau-level (LL) gaps. Near $D=0$ and for $n\geq0.7\times10^{12}~\mathrm{cm^{-2}}$, as guided by the white dashed lines in Fig.~\ref{fig4}(a), 
the LL sequence evolves with $\delta\nu=4$, where $\nu=nh/eB$. This sequence originates from the multiband R6 region, which contains light- and heavy-band Fermi pockets, each with fourfold spin and valley degeneracy. The switching between the light- and heavy-band LLs is shown in the enlarged map in Sec.~IX of the SM~\cite{SM}, consistent with Ref.~\cite{PhysRevLett.120.096802}. In contrast, in the single-pocket R5 region, guided by the yellow dashed lines in Fig.~\ref{fig4}(a), 
the $\delta\nu$ changes from 4 at smaller $D$ to 2 as $D$ increases and approaches the R4 boundary, indicating valley degeneracy lifting.

The valley splitting with increasing $D$ becomes even more evident at lower densities ($n\leq0.5\times10^{12}~\mathrm{cm^{-2}}$). At $D=0$, the first few LL sequences occur at $\nu=8,12,16,$ and $20$, whereas for $D/\epsilon_0\geq0.1~\mathrm{V\,nm^{-1}}$, the sequence evolves to $\nu=6,12,$ and $18$ [Fig.~\ref{fig4}(a)]. Figures~\ref{fig4}(c) and (d) show line cuts of $R_{xx}$ and $R_{xy}$ at $D/\epsilon_0=0$ and $0.5~\mathrm{V\,nm^{-1}}$, respectively. The $\nu=12$ state persists over the entire displacement-field range. In contrast, the $\nu=6$ state emerges only beyond a critical $D$, and its $R_{xx}$ minimum strengthens with increasing $D$, becoming more pronounced than that at $\nu=12$ [Fig.~\ref{fig4}(d)]. The emergence of $\nu=6$ from the threefold Dirac-gully Fermi surface (R1), with a total degeneracy of 12, further confirms the lifting of valley degeneracy.

To determine the origin of the splitting, we calculate the single-particle LL spectrum of B-4LG as a function of energy and $D$ at $B_\perp=1~\mathrm{T}$ [Fig.~\ref{fig4}(b)]. The red and blue lines denote LLs from the $K$ and $K'$ valleys, respectively, while each LL remains spin degenerate. The calculation shows a complete lifting of the valley degeneracy of the lowest LLs at finite $D$, with a gap of about $6~\mathrm{meV}$ at $D/\epsilon_0\sim0.6~\mathrm{V\,nm^{-1}}$. It also reproduces the valley splitting in the R5 region (black arrows in Fig.~\ref{fig4}(b)) and the sequence of incompressible states at $\nu=8,12,$ and $16$ observed at $D=0$ [Figs.~\ref{fig4}(b) and \ref{fig4}(c)]. Furthermore, the calculated LL crossings (dashed black box) agree well with experiment [Fig.~\ref{fig4}(a)], reflecting both valley splitting and crossings with LLs originating from other Fermi-surface regions (e.g., R2).

The large valley splitting reflects an enhanced valley-orbital response in B-4LG. The displacement field breaks inversion symmetry, polarizes the low-energy wave functions across the four layers, and hybridizes the nearby light- and heavy-mass bilayer-like bands. The resulting avoided crossings generate large valley-contrasting orbital magnetic moments with opposite signs. A perpendicular magnetic field couples to the orbital magnetic moments in the $K$ and $K'$ valleys, shifting their conduction-band edges in opposite directions and producing the large valley splitting calculated in Fig.~\ref{fig4}(e) and also observed experimentally (see \hyperref[sec:valley_orbital_mechanism]{EM2} for details). This mechanism explains the observed degeneracy lifting within the noninteracting quasiparticle picture, without invoking electron-electron interactions.

Previous studies have reported anomalous Hall responses near charge neutrality and attributed them to spontaneous valley symmetry breaking~\cite{chen2023gate}. Here, we observe no detectable anomalous Hall response in the hysteresis data (Sec.~XI of the SM~\cite{SM} for details). The finite-$D$ valley splitting in our experiment is quantitatively reproduced by the single-particle LL spectrum. The valley splitting gap of $\nu = 6$ at $B = 1\,T$ was further measured via an activation plot (see SM Sec.~XI~\cite{SM}), increasing from $\simeq2.0\,\text{meV}$ at $D=0.2\,V/nm$ to $\simeq4.5\,\text{meV}$ at $D=0.6\,V/nm$. This qualitatively agrees with the calculated single-particle LL spectrum. 

\tc{blue}{\textit{Conclusion:---}}
In summary, we have quantitatively reconstructed the evolution of Fermi-surface topology and flavor degeneracy across a displacement-field-tunable sequence of Lifshitz transitions in Bernal-stacked tetralayer graphene. Using Shubnikov--de Haas oscillation spectroscopy, we determined the extremal Fermi-surface areas together with their spin, valley, and gully degeneracies across six distinct Fermi-surface topologies, in excellent agreement with tight-binding calculations. Furthermore, the evolution of the incompressible quantum Hall states demonstrates that the combined action of displacement and magnetic fields lifts the valley degeneracy, giving rise to a large single-particle valley splitting. Together, these results establish quantum oscillation spectroscopy as a quantitative probe of both Fermi-surface topology and flavor degeneracy in multiband quantum materials.

\tc{blue}{\textit{Acknowledgement:---}}
H.V. acknowledges the Ministry of Education, Government of India, for financial support through the Prime Minister’s Research Fellowship. A.A. acknowledges funding from the Core Research Grant by ANRF (Sanction No. CRG/2023/007003), Department of Science and Technology, India. A.D. thanks the ANRF with project no: SP/ANRF-26-0244 and Department of Science and Technology (DST/NM/TUE/QM-5/2023) for the financial support.
\bibliography{ref}

\begin{thebibliography}{55}%
\makeatletter
\providecommand \@ifxundefined [1]{%
 \@ifx{#1\undefined}
}%
\providecommand \@ifnum [1]{%
 \ifnum #1\expandafter \@firstoftwo
 \else \expandafter \@secondoftwo
 \fi
}%
\providecommand \@ifx [1]{%
 \ifx #1\expandafter \@firstoftwo
 \else \expandafter \@secondoftwo
 \fi
}%
\providecommand \natexlab [1]{#1}%
\providecommand \enquote  [1]{``#1''}%
\providecommand \bibnamefont  [1]{#1}%
\providecommand \bibfnamefont [1]{#1}%
\providecommand \citenamefont [1]{#1}%
\providecommand \href@noop [0]{\@secondoftwo}%
\providecommand \href [0]{\begingroup \@sanitize@url \@href}%
\providecommand \@href[1]{\@@startlink{#1}\@@href}%
\providecommand \@@href[1]{\endgroup#1\@@endlink}%
\providecommand \@sanitize@url [0]{\catcode `\\12\catcode `\$12\catcode `\&12\catcode `\#12\catcode `\^12\catcode `\_12\catcode `\%12\relax}%
\providecommand \@@startlink[1]{}%
\providecommand \@@endlink[0]{}%
\providecommand \url  [0]{\begingroup\@sanitize@url \@url }%
\providecommand \@url [1]{\endgroup\@href {#1}{\urlprefix }}%
\providecommand \urlprefix  [0]{URL }%
\providecommand \Eprint [0]{\href }%
\providecommand \doibase [0]{https://doi.org/}%
\providecommand \selectlanguage [0]{\@gobble}%
\providecommand \bibinfo  [0]{\@secondoftwo}%
\providecommand \bibfield  [0]{\@secondoftwo}%
\providecommand \translation [1]{[#1]}%
\providecommand \BibitemOpen [0]{}%
\providecommand \bibitemStop [0]{}%
\providecommand \bibitemNoStop [0]{.\EOS\space}%
\providecommand \EOS [0]{\spacefactor3000\relax}%
\providecommand \BibitemShut  [1]{\csname bibitem#1\endcsname}%
\let\auto@bib@innerbib\@empty
\bibitem [{\citenamefont {Lifshitz}(1960)}]{lifshitz1960anomalies}%
  \BibitemOpen
  \bibfield  {author} {\bibinfo {author} {\bibfnamefont {I.~M.}\ \bibnamefont {Lifshitz}},\ }\bibfield  {title} {\bibinfo {title} {Anomalies of electron characteristics in the high pressure region},\ }\href {https://www.osti.gov/biblio/4173345} {\bibfield  {journal} {\bibinfo  {journal} {Zhur. Eksptl'. i Teoret. Fiz.}\ }\textbf {\bibinfo {volume} {Vol: 38}} (\bibinfo {year} {1960})}\BibitemShut {NoStop}%
\bibitem [{\citenamefont {Volovik}(2018)}]{Volovik_2018}%
  \BibitemOpen
  \bibfield  {author} {\bibinfo {author} {\bibfnamefont {G.~E.}\ \bibnamefont {Volovik}},\ }\bibfield  {title} {\bibinfo {title} {Exotic lifshitz transitions in topological materials},\ }\href {https://doi.org/10.3367/UFNe.2017.01.038218} {\bibfield  {journal} {\bibinfo  {journal} {Physics-Uspekhi}\ }\textbf {\bibinfo {volume} {61}},\ \bibinfo {pages} {89} (\bibinfo {year} {2018})}\BibitemShut {NoStop}%
\bibitem [{\citenamefont {Vignolle}\ \emph {et~al.}(2008)\citenamefont {Vignolle}, \citenamefont {Carrington}, \citenamefont {Cooper}, \citenamefont {French}, \citenamefont {Mackenzie}, \citenamefont {Jaudet}, \citenamefont {Vignolles}, \citenamefont {Proust},\ and\ \citenamefont {Hussey}}]{vignolle2008quantum}%
  \BibitemOpen
  \bibfield  {author} {\bibinfo {author} {\bibfnamefont {B.}~\bibnamefont {Vignolle}}, \bibinfo {author} {\bibfnamefont {A.}~\bibnamefont {Carrington}}, \bibinfo {author} {\bibfnamefont {R.}~\bibnamefont {Cooper}}, \bibinfo {author} {\bibfnamefont {M.}~\bibnamefont {French}}, \bibinfo {author} {\bibfnamefont {A.}~\bibnamefont {Mackenzie}}, \bibinfo {author} {\bibfnamefont {C.}~\bibnamefont {Jaudet}}, \bibinfo {author} {\bibfnamefont {D.}~\bibnamefont {Vignolles}}, \bibinfo {author} {\bibfnamefont {C.}~\bibnamefont {Proust}},\ and\ \bibinfo {author} {\bibfnamefont {N.}~\bibnamefont {Hussey}},\ }\bibfield  {title} {\bibinfo {title} {Quantum oscillations in an overdoped high-t c superconductor},\ }\href {https://doi.org/https://doi.org/10.1038/nature07323} {\bibfield  {journal} {\bibinfo  {journal} {Nature}\ }\textbf {\bibinfo {volume} {455}},\ \bibinfo {pages} {952} (\bibinfo {year} {2008})}\BibitemShut {NoStop}%
\bibitem [{\citenamefont {Benhabib}\ \emph {et~al.}(2015)\citenamefont {Benhabib}, \citenamefont {Sacuto}, \citenamefont {Civelli}, \citenamefont {Paul}, \citenamefont {Cazayous}, \citenamefont {Gallais}, \citenamefont {M\'easson}, \citenamefont {Zhong}, \citenamefont {Schneeloch}, \citenamefont {Gu}, \citenamefont {Colson},\ and\ \citenamefont {Forget}}]{PhysRevLett.114.147001}%
  \BibitemOpen
  \bibfield  {author} {\bibinfo {author} {\bibfnamefont {S.}~\bibnamefont {Benhabib}}, \bibinfo {author} {\bibfnamefont {A.}~\bibnamefont {Sacuto}}, \bibinfo {author} {\bibfnamefont {M.}~\bibnamefont {Civelli}}, \bibinfo {author} {\bibfnamefont {I.}~\bibnamefont {Paul}}, \bibinfo {author} {\bibfnamefont {M.}~\bibnamefont {Cazayous}}, \bibinfo {author} {\bibfnamefont {Y.}~\bibnamefont {Gallais}}, \bibinfo {author} {\bibfnamefont {M.-A.}\ \bibnamefont {M\'easson}}, \bibinfo {author} {\bibfnamefont {R.~D.}\ \bibnamefont {Zhong}}, \bibinfo {author} {\bibfnamefont {J.}~\bibnamefont {Schneeloch}}, \bibinfo {author} {\bibfnamefont {G.~D.}\ \bibnamefont {Gu}}, \bibinfo {author} {\bibfnamefont {D.}~\bibnamefont {Colson}},\ and\ \bibinfo {author} {\bibfnamefont {A.}~\bibnamefont {Forget}},\ }\bibfield  {title} {\bibinfo {title} {Collapse of the normal-state pseudogap at a lifshitz transition in the ${\mathrm{bi}}_{2}{\mathrm{sr}}_{2}{\mathrm{cacu}}_{2}{\mathrm{o}}_{8+\ensuremath{\delta}}$ cuprate superconductor},\
  }\href {https://doi.org/10.1103/PhysRevLett.114.147001} {\bibfield  {journal} {\bibinfo  {journal} {Phys. Rev. Lett.}\ }\textbf {\bibinfo {volume} {114}},\ \bibinfo {pages} {147001} (\bibinfo {year} {2015})}\BibitemShut {NoStop}%
\bibitem [{\citenamefont {Liu}\ \emph {et~al.}(2010)\citenamefont {Liu}, \citenamefont {Kondo}, \citenamefont {Fernandes}, \citenamefont {Palczewski}, \citenamefont {Mun}, \citenamefont {Ni}, \citenamefont {Thaler}, \citenamefont {Bostwick}, \citenamefont {Rotenberg}, \citenamefont {Schmalian} \emph {et~al.}}]{liu2010evidence}%
  \BibitemOpen
  \bibfield  {author} {\bibinfo {author} {\bibfnamefont {C.}~\bibnamefont {Liu}}, \bibinfo {author} {\bibfnamefont {T.}~\bibnamefont {Kondo}}, \bibinfo {author} {\bibfnamefont {R.~M.}\ \bibnamefont {Fernandes}}, \bibinfo {author} {\bibfnamefont {A.~D.}\ \bibnamefont {Palczewski}}, \bibinfo {author} {\bibfnamefont {E.~D.}\ \bibnamefont {Mun}}, \bibinfo {author} {\bibfnamefont {N.}~\bibnamefont {Ni}}, \bibinfo {author} {\bibfnamefont {A.~N.}\ \bibnamefont {Thaler}}, \bibinfo {author} {\bibfnamefont {A.}~\bibnamefont {Bostwick}}, \bibinfo {author} {\bibfnamefont {E.}~\bibnamefont {Rotenberg}}, \bibinfo {author} {\bibfnamefont {J.}~\bibnamefont {Schmalian}}, \emph {et~al.},\ }\bibfield  {title} {\bibinfo {title} {Evidence for a lifshitz transition in electron-doped iron arsenic superconductors at the onset of superconductivity},\ }\href {https://doi.org/https://doi.org/10.1038/nphys1656} {\bibfield  {journal} {\bibinfo  {journal} {Nature Physics}\ }\textbf {\bibinfo {volume} {6}},\ \bibinfo {pages} {419}
  (\bibinfo {year} {2010})}\BibitemShut {NoStop}%
\bibitem [{\citenamefont {Ren}\ \emph {et~al.}(2017)\citenamefont {Ren}, \citenamefont {Yan}, \citenamefont {Niu}, \citenamefont {Tao}, \citenamefont {Hu}, \citenamefont {Peng}, \citenamefont {Xie}, \citenamefont {Zhao}, \citenamefont {Zhang},\ and\ \citenamefont {Feng}}]{Ren2017SciAdv}%
  \BibitemOpen
  \bibfield  {author} {\bibinfo {author} {\bibfnamefont {M.}~\bibnamefont {Ren}}, \bibinfo {author} {\bibfnamefont {Y.}~\bibnamefont {Yan}}, \bibinfo {author} {\bibfnamefont {X.}~\bibnamefont {Niu}}, \bibinfo {author} {\bibfnamefont {R.}~\bibnamefont {Tao}}, \bibinfo {author} {\bibfnamefont {D.}~\bibnamefont {Hu}}, \bibinfo {author} {\bibfnamefont {R.}~\bibnamefont {Peng}}, \bibinfo {author} {\bibfnamefont {B.}~\bibnamefont {Xie}}, \bibinfo {author} {\bibfnamefont {J.}~\bibnamefont {Zhao}}, \bibinfo {author} {\bibfnamefont {T.}~\bibnamefont {Zhang}},\ and\ \bibinfo {author} {\bibfnamefont {D.-L.}\ \bibnamefont {Feng}},\ }\bibfield  {title} {\bibinfo {title} {Superconductivity across lifshitz transition and anomalous insulating state in surface {K}-dosed {($\mathrm{Li}_{0.8}\mathrm{Fe}_{0.2}\mathrm{OH}$)FeSe}},\ }\href {https://doi.org/10.1126/sciadv.1603238} {\bibfield  {journal} {\bibinfo  {journal} {Science Advances}\ }\textbf {\bibinfo {volume} {3}},\ \bibinfo {pages} {e1603238} (\bibinfo {year}
  {2017})}\BibitemShut {NoStop}%
\bibitem [{\citenamefont {Yelland}\ \emph {et~al.}(2011)\citenamefont {Yelland}, \citenamefont {Barraclough}, \citenamefont {Wang}, \citenamefont {Kamenev},\ and\ \citenamefont {Huxley}}]{yelland2011high}%
  \BibitemOpen
  \bibfield  {author} {\bibinfo {author} {\bibfnamefont {E.~A.}\ \bibnamefont {Yelland}}, \bibinfo {author} {\bibfnamefont {J.~M.}\ \bibnamefont {Barraclough}}, \bibinfo {author} {\bibfnamefont {W.}~\bibnamefont {Wang}}, \bibinfo {author} {\bibfnamefont {K.}~\bibnamefont {Kamenev}},\ and\ \bibinfo {author} {\bibfnamefont {A.~D.}\ \bibnamefont {Huxley}},\ }\bibfield  {title} {\bibinfo {title} {High-field superconductivity at an electronic topological transition in urhge},\ }\href {https://doi.org/https://doi.org/10.1038/nphys2073} {\bibfield  {journal} {\bibinfo  {journal} {Nature physics}\ }\textbf {\bibinfo {volume} {7}},\ \bibinfo {pages} {890} (\bibinfo {year} {2011})}\BibitemShut {NoStop}%
\bibitem [{\citenamefont {Pfau}\ \emph {et~al.}(2013)\citenamefont {Pfau}, \citenamefont {Daou}, \citenamefont {Lausberg}, \citenamefont {Naren}, \citenamefont {Brando}, \citenamefont {Friedemann}, \citenamefont {Wirth}, \citenamefont {Westerkamp}, \citenamefont {Stockert}, \citenamefont {Gegenwart}, \citenamefont {Krellner}, \citenamefont {Geibel}, \citenamefont {Zwicknagl},\ and\ \citenamefont {Steglich}}]{PhysRevLett.110.256403}%
  \BibitemOpen
  \bibfield  {author} {\bibinfo {author} {\bibfnamefont {H.}~\bibnamefont {Pfau}}, \bibinfo {author} {\bibfnamefont {R.}~\bibnamefont {Daou}}, \bibinfo {author} {\bibfnamefont {S.}~\bibnamefont {Lausberg}}, \bibinfo {author} {\bibfnamefont {H.~R.}\ \bibnamefont {Naren}}, \bibinfo {author} {\bibfnamefont {M.}~\bibnamefont {Brando}}, \bibinfo {author} {\bibfnamefont {S.}~\bibnamefont {Friedemann}}, \bibinfo {author} {\bibfnamefont {S.}~\bibnamefont {Wirth}}, \bibinfo {author} {\bibfnamefont {T.}~\bibnamefont {Westerkamp}}, \bibinfo {author} {\bibfnamefont {U.}~\bibnamefont {Stockert}}, \bibinfo {author} {\bibfnamefont {P.}~\bibnamefont {Gegenwart}}, \bibinfo {author} {\bibfnamefont {C.}~\bibnamefont {Krellner}}, \bibinfo {author} {\bibfnamefont {C.}~\bibnamefont {Geibel}}, \bibinfo {author} {\bibfnamefont {G.}~\bibnamefont {Zwicknagl}},\ and\ \bibinfo {author} {\bibfnamefont {F.}~\bibnamefont {Steglich}},\ }\bibfield  {title} {\bibinfo {title} {Interplay between kondo suppression and lifshitz transitions in
  ${\mathrm{ybrh}}_{2}{\mathrm{si}}_{2}$ at high magnetic fields},\ }\href {https://doi.org/10.1103/PhysRevLett.110.256403} {\bibfield  {journal} {\bibinfo  {journal} {Phys. Rev. Lett.}\ }\textbf {\bibinfo {volume} {110}},\ \bibinfo {pages} {256403} (\bibinfo {year} {2013})}\BibitemShut {NoStop}%
\bibitem [{\citenamefont {Niu}\ \emph {et~al.}(2020)\citenamefont {Niu}, \citenamefont {Knebel}, \citenamefont {Braithwaite}, \citenamefont {Aoki}, \citenamefont {Lapertot}, \citenamefont {Seyfarth}, \citenamefont {Brison}, \citenamefont {Flouquet},\ and\ \citenamefont {Pourret}}]{PhysRevLett.124.086601}%
  \BibitemOpen
  \bibfield  {author} {\bibinfo {author} {\bibfnamefont {Q.}~\bibnamefont {Niu}}, \bibinfo {author} {\bibfnamefont {G.}~\bibnamefont {Knebel}}, \bibinfo {author} {\bibfnamefont {D.}~\bibnamefont {Braithwaite}}, \bibinfo {author} {\bibfnamefont {D.}~\bibnamefont {Aoki}}, \bibinfo {author} {\bibfnamefont {G.}~\bibnamefont {Lapertot}}, \bibinfo {author} {\bibfnamefont {G.}~\bibnamefont {Seyfarth}}, \bibinfo {author} {\bibfnamefont {J.-P.}\ \bibnamefont {Brison}}, \bibinfo {author} {\bibfnamefont {J.}~\bibnamefont {Flouquet}},\ and\ \bibinfo {author} {\bibfnamefont {A.}~\bibnamefont {Pourret}},\ }\bibfield  {title} {\bibinfo {title} {Fermi-surface instability in the heavy-fermion superconductor ${\mathrm{ute}}_{2}$},\ }\href {https://doi.org/10.1103/PhysRevLett.124.086601} {\bibfield  {journal} {\bibinfo  {journal} {Phys. Rev. Lett.}\ }\textbf {\bibinfo {volume} {124}},\ \bibinfo {pages} {086601} (\bibinfo {year} {2020})}\BibitemShut {NoStop}%
\bibitem [{\citenamefont {Yang}\ \emph {et~al.}(2019)\citenamefont {Yang}, \citenamefont {Yang}, \citenamefont {Liu}, \citenamefont {Sun}, \citenamefont {Chen}, \citenamefont {Peng}, \citenamefont {Schmidt}, \citenamefont {Prabhakaran}, \citenamefont {Bernevig}, \citenamefont {Felser} \emph {et~al.}}]{yang2019topological}%
  \BibitemOpen
  \bibfield  {author} {\bibinfo {author} {\bibfnamefont {H.}~\bibnamefont {Yang}}, \bibinfo {author} {\bibfnamefont {L.}~\bibnamefont {Yang}}, \bibinfo {author} {\bibfnamefont {Z.}~\bibnamefont {Liu}}, \bibinfo {author} {\bibfnamefont {Y.}~\bibnamefont {Sun}}, \bibinfo {author} {\bibfnamefont {C.}~\bibnamefont {Chen}}, \bibinfo {author} {\bibfnamefont {H.}~\bibnamefont {Peng}}, \bibinfo {author} {\bibfnamefont {M.}~\bibnamefont {Schmidt}}, \bibinfo {author} {\bibfnamefont {D.}~\bibnamefont {Prabhakaran}}, \bibinfo {author} {\bibfnamefont {B.~A.}\ \bibnamefont {Bernevig}}, \bibinfo {author} {\bibfnamefont {C.}~\bibnamefont {Felser}}, \emph {et~al.},\ }\bibfield  {title} {\bibinfo {title} {Topological lifshitz transitions and fermi arc manipulation in weyl semimetal nbas},\ }\href {https://doi.org/https://doi.org/10.1038/s41467-019-11491-4} {\bibfield  {journal} {\bibinfo  {journal} {Nature communications}\ }\textbf {\bibinfo {volume} {10}},\ \bibinfo {pages} {3478} (\bibinfo {year} {2019})}\BibitemShut
  {NoStop}%
\bibitem [{\citenamefont {Liu}\ \emph {et~al.}(2020)\citenamefont {Liu}, \citenamefont {Liu}, \citenamefont {Gui}, \citenamefont {Xiang}, \citenamefont {Zhou}, \citenamefont {Hsu}, \citenamefont {Lin}, \citenamefont {Chang}, \citenamefont {Xie},\ and\ \citenamefont {Jia}}]{Liu2020PNAS}%
  \BibitemOpen
  \bibfield  {author} {\bibinfo {author} {\bibfnamefont {Y.}~\bibnamefont {Liu}}, \bibinfo {author} {\bibfnamefont {Y.-F.}\ \bibnamefont {Liu}}, \bibinfo {author} {\bibfnamefont {X.}~\bibnamefont {Gui}}, \bibinfo {author} {\bibfnamefont {C.}~\bibnamefont {Xiang}}, \bibinfo {author} {\bibfnamefont {H.-B.}\ \bibnamefont {Zhou}}, \bibinfo {author} {\bibfnamefont {C.-H.}\ \bibnamefont {Hsu}}, \bibinfo {author} {\bibfnamefont {H.}~\bibnamefont {Lin}}, \bibinfo {author} {\bibfnamefont {T.-R.}\ \bibnamefont {Chang}}, \bibinfo {author} {\bibfnamefont {W.}~\bibnamefont {Xie}},\ and\ \bibinfo {author} {\bibfnamefont {S.}~\bibnamefont {Jia}},\ }\bibfield  {title} {\bibinfo {title} {Bond-breaking induced lifshitz transition in robust dirac semimetal {VAl$_3$}},\ }\href {https://doi.org/10.1073/pnas.1917697117} {\bibfield  {journal} {\bibinfo  {journal} {Proceedings of the National Academy of Sciences}\ }\textbf {\bibinfo {volume} {117}},\ \bibinfo {pages} {15517} (\bibinfo {year} {2020})}\BibitemShut {NoStop}%
\bibitem [{\citenamefont {Shi}\ \emph {et~al.}(2017)\citenamefont {Shi}, \citenamefont {Han}, \citenamefont {Peng}, \citenamefont {Richard}, \citenamefont {Qian}, \citenamefont {Wu}, \citenamefont {Qiu}, \citenamefont {Wang}, \citenamefont {Hu}, \citenamefont {Sun} \emph {et~al.}}]{shi2017enhanced}%
  \BibitemOpen
  \bibfield  {author} {\bibinfo {author} {\bibfnamefont {X.}~\bibnamefont {Shi}}, \bibinfo {author} {\bibfnamefont {Z.}~\bibnamefont {Han}}, \bibinfo {author} {\bibfnamefont {X.}~\bibnamefont {Peng}}, \bibinfo {author} {\bibfnamefont {P.}~\bibnamefont {Richard}}, \bibinfo {author} {\bibfnamefont {T.}~\bibnamefont {Qian}}, \bibinfo {author} {\bibfnamefont {X.}~\bibnamefont {Wu}}, \bibinfo {author} {\bibfnamefont {M.}~\bibnamefont {Qiu}}, \bibinfo {author} {\bibfnamefont {S.}~\bibnamefont {Wang}}, \bibinfo {author} {\bibfnamefont {J.}~\bibnamefont {Hu}}, \bibinfo {author} {\bibfnamefont {Y.}~\bibnamefont {Sun}}, \emph {et~al.},\ }\bibfield  {title} {\bibinfo {title} {Enhanced superconductivity accompanying a lifshitz transition in electron-doped fese monolayer},\ }\href {https://doi.org/https://doi.org/10.1038/ncomms14988} {\bibfield  {journal} {\bibinfo  {journal} {Nature communications}\ }\textbf {\bibinfo {volume} {8}},\ \bibinfo {pages} {14988} (\bibinfo {year} {2017})}\BibitemShut {NoStop}%
\bibitem [{\citenamefont {Varlet}\ \emph {et~al.}(2015)\citenamefont {Varlet}, \citenamefont {Mucha-Kruczyński}, \citenamefont {Bischoff}, \citenamefont {Simonet}, \citenamefont {Taniguchi}, \citenamefont {Watanabe}, \citenamefont {Fal’ko}, \citenamefont {Ihn},\ and\ \citenamefont {Ensslin}}]{VARLET201519}%
  \BibitemOpen
  \bibfield  {author} {\bibinfo {author} {\bibfnamefont {A.}~\bibnamefont {Varlet}}, \bibinfo {author} {\bibfnamefont {M.}~\bibnamefont {Mucha-Kruczyński}}, \bibinfo {author} {\bibfnamefont {D.}~\bibnamefont {Bischoff}}, \bibinfo {author} {\bibfnamefont {P.}~\bibnamefont {Simonet}}, \bibinfo {author} {\bibfnamefont {T.}~\bibnamefont {Taniguchi}}, \bibinfo {author} {\bibfnamefont {K.}~\bibnamefont {Watanabe}}, \bibinfo {author} {\bibfnamefont {V.}~\bibnamefont {Fal’ko}}, \bibinfo {author} {\bibfnamefont {T.}~\bibnamefont {Ihn}},\ and\ \bibinfo {author} {\bibfnamefont {K.}~\bibnamefont {Ensslin}},\ }\bibfield  {title} {\bibinfo {title} {Tunable fermi surface topology and lifshitz transition in bilayer graphene},\ }\href {https://doi.org/https://doi.org/10.1016/j.synthmet.2015.07.006} {\bibfield  {journal} {\bibinfo  {journal} {Synthetic Metals}\ }\textbf {\bibinfo {volume} {210}},\ \bibinfo {pages} {19} (\bibinfo {year} {2015})}\BibitemShut {NoStop}%
\bibitem [{\citenamefont {Shi}\ \emph {et~al.}(2018)\citenamefont {Shi}, \citenamefont {Che}, \citenamefont {Zhou}, \citenamefont {Ge}, \citenamefont {Pi}, \citenamefont {Espiritu}, \citenamefont {Taniguchi}, \citenamefont {Watanabe}, \citenamefont {Barlas}, \citenamefont {Lake},\ and\ \citenamefont {Lau}}]{PhysRevLett.120.096802}%
  \BibitemOpen
  \bibfield  {author} {\bibinfo {author} {\bibfnamefont {Y.}~\bibnamefont {Shi}}, \bibinfo {author} {\bibfnamefont {S.}~\bibnamefont {Che}}, \bibinfo {author} {\bibfnamefont {K.}~\bibnamefont {Zhou}}, \bibinfo {author} {\bibfnamefont {S.}~\bibnamefont {Ge}}, \bibinfo {author} {\bibfnamefont {Z.}~\bibnamefont {Pi}}, \bibinfo {author} {\bibfnamefont {T.}~\bibnamefont {Espiritu}}, \bibinfo {author} {\bibfnamefont {T.}~\bibnamefont {Taniguchi}}, \bibinfo {author} {\bibfnamefont {K.}~\bibnamefont {Watanabe}}, \bibinfo {author} {\bibfnamefont {Y.}~\bibnamefont {Barlas}}, \bibinfo {author} {\bibfnamefont {R.}~\bibnamefont {Lake}},\ and\ \bibinfo {author} {\bibfnamefont {C.~N.}\ \bibnamefont {Lau}},\ }\bibfield  {title} {\bibinfo {title} {Tunable lifshitz transitions and multiband transport in tetralayer graphene},\ }\href {https://doi.org/10.1103/PhysRevLett.120.096802} {\bibfield  {journal} {\bibinfo  {journal} {Phys. Rev. Lett.}\ }\textbf {\bibinfo {volume} {120}},\ \bibinfo {pages} {096802} (\bibinfo {year}
  {2018})}\BibitemShut {NoStop}%
\bibitem [{\citenamefont {Jayaraman}\ \emph {et~al.}(2021)\citenamefont {Jayaraman}, \citenamefont {Hsieh}, \citenamefont {Ghawri}, \citenamefont {Mahapatra}, \citenamefont {Watanabe}, \citenamefont {Taniguchi},\ and\ \citenamefont {Ghosh}}]{jayaraman2021evidence}%
  \BibitemOpen
  \bibfield  {author} {\bibinfo {author} {\bibfnamefont {A.}~\bibnamefont {Jayaraman}}, \bibinfo {author} {\bibfnamefont {K.}~\bibnamefont {Hsieh}}, \bibinfo {author} {\bibfnamefont {B.}~\bibnamefont {Ghawri}}, \bibinfo {author} {\bibfnamefont {P.~S.}\ \bibnamefont {Mahapatra}}, \bibinfo {author} {\bibfnamefont {K.}~\bibnamefont {Watanabe}}, \bibinfo {author} {\bibfnamefont {T.}~\bibnamefont {Taniguchi}},\ and\ \bibinfo {author} {\bibfnamefont {A.}~\bibnamefont {Ghosh}},\ }\bibfield  {title} {\bibinfo {title} {Evidence of lifshitz transition in the thermoelectric power of ultrahigh-mobility bilayer graphene},\ }\href@noop {} {\bibfield  {journal} {\bibinfo  {journal} {Nano Letters}\ }\textbf {\bibinfo {volume} {21}},\ \bibinfo {pages} {1221} (\bibinfo {year} {2021})}\BibitemShut {NoStop}%
\bibitem [{\citenamefont {Datta}\ \emph {et~al.}(2024)\citenamefont {Datta}, \citenamefont {Bhowmik}, \citenamefont {Varshney}, \citenamefont {Watanabe}, \citenamefont {Taniguchi}, \citenamefont {Agarwal},\ and\ \citenamefont {Chandni}}]{Datta2024nonlinear}%
  \BibitemOpen
  \bibfield  {author} {\bibinfo {author} {\bibfnamefont {S.}~\bibnamefont {Datta}}, \bibinfo {author} {\bibfnamefont {S.}~\bibnamefont {Bhowmik}}, \bibinfo {author} {\bibfnamefont {H.}~\bibnamefont {Varshney}}, \bibinfo {author} {\bibfnamefont {K.}~\bibnamefont {Watanabe}}, \bibinfo {author} {\bibfnamefont {T.}~\bibnamefont {Taniguchi}}, \bibinfo {author} {\bibfnamefont {A.}~\bibnamefont {Agarwal}},\ and\ \bibinfo {author} {\bibfnamefont {U.}~\bibnamefont {Chandni}},\ }\bibfield  {title} {\bibinfo {title} {Nonlinear electrical transport unveils fermi surface malleability in a moiré heterostructure},\ }\href {https://doi.org/10.1021/acs.nanolett.4c01946} {\bibfield  {journal} {\bibinfo  {journal} {Nano Letters}\ }\textbf {\bibinfo {volume} {24}},\ \bibinfo {pages} {9520–9527} (\bibinfo {year} {2024})}\BibitemShut {NoStop}%
\bibitem [{\citenamefont {Ahmed}\ \emph {et~al.}(2025)\citenamefont {Ahmed}, \citenamefont {Varshney}, \citenamefont {Tu}, \citenamefont {Watanabe}, \citenamefont {Taniguchi}, \citenamefont {Gobbi}, \citenamefont {Casanova}, \citenamefont {Agarwal},\ and\ \citenamefont {Hueso}}]{Ahmed2025}%
  \BibitemOpen
  \bibfield  {author} {\bibinfo {author} {\bibfnamefont {T.}~\bibnamefont {Ahmed}}, \bibinfo {author} {\bibfnamefont {H.}~\bibnamefont {Varshney}}, \bibinfo {author} {\bibfnamefont {B.~Q.}\ \bibnamefont {Tu}}, \bibinfo {author} {\bibfnamefont {K.}~\bibnamefont {Watanabe}}, \bibinfo {author} {\bibfnamefont {T.}~\bibnamefont {Taniguchi}}, \bibinfo {author} {\bibfnamefont {M.}~\bibnamefont {Gobbi}}, \bibinfo {author} {\bibfnamefont {F.}~\bibnamefont {Casanova}}, \bibinfo {author} {\bibfnamefont {A.}~\bibnamefont {Agarwal}},\ and\ \bibinfo {author} {\bibfnamefont {L.~E.}\ \bibnamefont {Hueso}},\ }\bibfield  {title} {\bibinfo {title} {Detecting lifshitz transitions using nonlinear conductivity in bilayer graphene},\ }\href {https://doi.org/10.1002/smll.202501426} {\bibfield  {journal} {\bibinfo  {journal} {Small}\ }\textbf {\bibinfo {volume} {21}} (\bibinfo {year} {2025})}\BibitemShut {NoStop}%
\bibitem [{\citenamefont {Sahani}\ \emph {et~al.}(2025)\citenamefont {Sahani}, \citenamefont {Das}, \citenamefont {Watanabe}, \citenamefont {Taniguchi}, \citenamefont {Agarwal},\ and\ \citenamefont {Bid}}]{Sahani2025OMR}%
  \BibitemOpen
  \bibfield  {author} {\bibinfo {author} {\bibfnamefont {D.}~\bibnamefont {Sahani}}, \bibinfo {author} {\bibfnamefont {S.}~\bibnamefont {Das}}, \bibinfo {author} {\bibfnamefont {K.}~\bibnamefont {Watanabe}}, \bibinfo {author} {\bibfnamefont {T.}~\bibnamefont {Taniguchi}}, \bibinfo {author} {\bibfnamefont {A.}~\bibnamefont {Agarwal}},\ and\ \bibinfo {author} {\bibfnamefont {A.}~\bibnamefont {Bid}},\ }\bibfield  {title} {\bibinfo {title} {Giant gate-controlled room-temperature odd-parity magnetoresistance in magnetized bilayer graphene},\ }\href {https://doi.org/10.1103/PhysRevLett.134.106301} {\bibfield  {journal} {\bibinfo  {journal} {Phys. Rev. Lett.}\ }\textbf {\bibinfo {volume} {134}},\ \bibinfo {pages} {106301} (\bibinfo {year} {2025})}\BibitemShut {NoStop}%
\bibitem [{\citenamefont {Castro~Neto}\ \emph {et~al.}(2009)\citenamefont {Castro~Neto}, \citenamefont {Guinea}, \citenamefont {Peres}, \citenamefont {Novoselov},\ and\ \citenamefont {Geim}}]{castro2009electronic}%
  \BibitemOpen
  \bibfield  {author} {\bibinfo {author} {\bibfnamefont {A.~H.}\ \bibnamefont {Castro~Neto}}, \bibinfo {author} {\bibfnamefont {F.}~\bibnamefont {Guinea}}, \bibinfo {author} {\bibfnamefont {N.~M.~R.}\ \bibnamefont {Peres}}, \bibinfo {author} {\bibfnamefont {K.~S.}\ \bibnamefont {Novoselov}},\ and\ \bibinfo {author} {\bibfnamefont {A.~K.}\ \bibnamefont {Geim}},\ }\bibfield  {title} {\bibinfo {title} {The electronic properties of graphene},\ }\href {https://doi.org/10.1103/RevModPhys.81.109} {\bibfield  {journal} {\bibinfo  {journal} {Rev. Mod. Phys.}\ }\textbf {\bibinfo {volume} {81}},\ \bibinfo {pages} {109} (\bibinfo {year} {2009})}\BibitemShut {NoStop}%
\bibitem [{\citenamefont {McCann}\ and\ \citenamefont {Koshino}(2013)}]{mccann2013electronic}%
  \BibitemOpen
  \bibfield  {author} {\bibinfo {author} {\bibfnamefont {E.}~\bibnamefont {McCann}}\ and\ \bibinfo {author} {\bibfnamefont {M.}~\bibnamefont {Koshino}},\ }\bibfield  {title} {\bibinfo {title} {The electronic properties of bilayer graphene},\ }\href {https://doi.org/10.1088/0034-4885/76/5/056503} {\bibfield  {journal} {\bibinfo  {journal} {Reports on Progress in physics}\ }\textbf {\bibinfo {volume} {76}},\ \bibinfo {pages} {056503} (\bibinfo {year} {2013})}\BibitemShut {NoStop}%
\bibitem [{\citenamefont {Avetisyan}\ \emph {et~al.}(2009{\natexlab{a}})\citenamefont {Avetisyan}, \citenamefont {Partoens},\ and\ \citenamefont {Peeters}}]{PhysRevB.79.035421}%
  \BibitemOpen
  \bibfield  {author} {\bibinfo {author} {\bibfnamefont {A.~A.}\ \bibnamefont {Avetisyan}}, \bibinfo {author} {\bibfnamefont {B.}~\bibnamefont {Partoens}},\ and\ \bibinfo {author} {\bibfnamefont {F.~M.}\ \bibnamefont {Peeters}},\ }\bibfield  {title} {\bibinfo {title} {Electric field tuning of the band gap in graphene multilayers},\ }\href {https://doi.org/10.1103/PhysRevB.79.035421} {\bibfield  {journal} {\bibinfo  {journal} {Phys. Rev. B}\ }\textbf {\bibinfo {volume} {79}},\ \bibinfo {pages} {035421} (\bibinfo {year} {2009}{\natexlab{a}})}\BibitemShut {NoStop}%
\bibitem [{\citenamefont {Klanurak}\ \emph {et~al.}(2024)\citenamefont {Klanurak}, \citenamefont {Watanabe}, \citenamefont {Taniguchi}, \citenamefont {Chatraphorn},\ and\ \citenamefont {Taychatanapat}}]{doi:10.1021/acs.nanolett.4c01133}%
  \BibitemOpen
  \bibfield  {author} {\bibinfo {author} {\bibfnamefont {I.}~\bibnamefont {Klanurak}}, \bibinfo {author} {\bibfnamefont {K.}~\bibnamefont {Watanabe}}, \bibinfo {author} {\bibfnamefont {T.}~\bibnamefont {Taniguchi}}, \bibinfo {author} {\bibfnamefont {S.}~\bibnamefont {Chatraphorn}},\ and\ \bibinfo {author} {\bibfnamefont {T.}~\bibnamefont {Taychatanapat}},\ }\bibfield  {title} {\bibinfo {title} {Probing the anisotropic fermi surface in tetralayer graphene via transverse magnetic focusing},\ }\href {https://doi.org/10.1021/acs.nanolett.4c01133} {\bibfield  {journal} {\bibinfo  {journal} {Nano Letters}\ }\textbf {\bibinfo {volume} {24}},\ \bibinfo {pages} {6330} (\bibinfo {year} {2024})}\BibitemShut {NoStop}%
\bibitem [{\citenamefont {Varshney}\ and\ \citenamefont {Agarwal}(2026)}]{Harsh2026asymmetric}%
  \BibitemOpen
  \bibfield  {author} {\bibinfo {author} {\bibfnamefont {H.}~\bibnamefont {Varshney}}\ and\ \bibinfo {author} {\bibfnamefont {A.}~\bibnamefont {Agarwal}},\ }\bibfield  {title} {\bibinfo {title} {Asymmetric scattering drives large nonlinear nernst and seebeck effects},\ }\href {https://doi.org/10.1103/973q-b957} {\bibfield  {journal} {\bibinfo  {journal} {Phys. Rev. B}\ }\textbf {\bibinfo {volume} {113}},\ \bibinfo {pages} {195417} (\bibinfo {year} {2026})}\BibitemShut {NoStop}%
\bibitem [{\citenamefont {Rao}\ and\ \citenamefont {Serbyn}(2020)}]{PhysRevB.101.245411}%
  \BibitemOpen
  \bibfield  {author} {\bibinfo {author} {\bibfnamefont {P.}~\bibnamefont {Rao}}\ and\ \bibinfo {author} {\bibfnamefont {M.}~\bibnamefont {Serbyn}},\ }\bibfield  {title} {\bibinfo {title} {Gully quantum hall ferromagnetism in biased trilayer graphene},\ }\href {https://doi.org/10.1103/PhysRevB.101.245411} {\bibfield  {journal} {\bibinfo  {journal} {Phys. Rev. B}\ }\textbf {\bibinfo {volume} {101}},\ \bibinfo {pages} {245411} (\bibinfo {year} {2020})}\BibitemShut {NoStop}%
\bibitem [{\citenamefont {Zibrov}\ \emph {et~al.}(2018)\citenamefont {Zibrov}, \citenamefont {Rao}, \citenamefont {Kometter}, \citenamefont {Spanton}, \citenamefont {Li}, \citenamefont {Dean}, \citenamefont {Taniguchi}, \citenamefont {Watanabe}, \citenamefont {Serbyn},\ and\ \citenamefont {Young}}]{PhysRevLett.121.167601}%
  \BibitemOpen
  \bibfield  {author} {\bibinfo {author} {\bibfnamefont {A.~A.}\ \bibnamefont {Zibrov}}, \bibinfo {author} {\bibfnamefont {P.}~\bibnamefont {Rao}}, \bibinfo {author} {\bibfnamefont {C.}~\bibnamefont {Kometter}}, \bibinfo {author} {\bibfnamefont {E.~M.}\ \bibnamefont {Spanton}}, \bibinfo {author} {\bibfnamefont {J.~I.~A.}\ \bibnamefont {Li}}, \bibinfo {author} {\bibfnamefont {C.~R.}\ \bibnamefont {Dean}}, \bibinfo {author} {\bibfnamefont {T.}~\bibnamefont {Taniguchi}}, \bibinfo {author} {\bibfnamefont {K.}~\bibnamefont {Watanabe}}, \bibinfo {author} {\bibfnamefont {M.}~\bibnamefont {Serbyn}},\ and\ \bibinfo {author} {\bibfnamefont {A.~F.}\ \bibnamefont {Young}},\ }\bibfield  {title} {\bibinfo {title} {Emergent dirac gullies and gully-symmetry-breaking quantum hall states in $aba$ trilayer graphene},\ }\href {https://doi.org/10.1103/PhysRevLett.121.167601} {\bibfield  {journal} {\bibinfo  {journal} {Phys. Rev. Lett.}\ }\textbf {\bibinfo {volume} {121}},\ \bibinfo {pages} {167601} (\bibinfo {year}
  {2018})}\BibitemShut {NoStop}%
\bibitem [{\citenamefont {Kaur}\ \emph {et~al.}(2025)\citenamefont {Kaur}, \citenamefont {Ghorai}, \citenamefont {Samanta}, \citenamefont {Watanabe}, \citenamefont {Taniguchi}, \citenamefont {Sensarma},\ and\ \citenamefont {Bid}}]{nm8b-5vgm}%
  \BibitemOpen
  \bibfield  {author} {\bibinfo {author} {\bibfnamefont {S.}~\bibnamefont {Kaur}}, \bibinfo {author} {\bibfnamefont {U.}~\bibnamefont {Ghorai}}, \bibinfo {author} {\bibfnamefont {A.}~\bibnamefont {Samanta}}, \bibinfo {author} {\bibfnamefont {K.}~\bibnamefont {Watanabe}}, \bibinfo {author} {\bibfnamefont {T.}~\bibnamefont {Taniguchi}}, \bibinfo {author} {\bibfnamefont {R.}~\bibnamefont {Sensarma}},\ and\ \bibinfo {author} {\bibfnamefont {A.}~\bibnamefont {Bid}},\ }\bibfield  {title} {\bibinfo {title} {Symmetry broken states at high displacement fields in aba trilayer graphene},\ }\href {https://doi.org/10.1103/nm8b-5vgm} {\bibfield  {journal} {\bibinfo  {journal} {Phys. Rev. B}\ }\textbf {\bibinfo {volume} {112}},\ \bibinfo {pages} {L161103} (\bibinfo {year} {2025})}\BibitemShut {NoStop}%
\bibitem [{\citenamefont {Qu}\ \emph {et~al.}(2025{\natexlab{a}})\citenamefont {Qu}, \citenamefont {Chen}, \citenamefont {Han}, \citenamefont {Wang}, \citenamefont {Li}, \citenamefont {Liu}, \citenamefont {Zhao}, \citenamefont {Watanabe}, \citenamefont {Taniguchi}, \citenamefont {Cheng}, \citenamefont {Gan},\ and\ \citenamefont {Lu}}]{Qu_2025}%
  \BibitemOpen
  \bibfield  {author} {\bibinfo {author} {\bibfnamefont {Z.}~\bibnamefont {Qu}}, \bibinfo {author} {\bibfnamefont {Z.}~\bibnamefont {Chen}}, \bibinfo {author} {\bibfnamefont {X.}~\bibnamefont {Han}}, \bibinfo {author} {\bibfnamefont {Z.}~\bibnamefont {Wang}}, \bibinfo {author} {\bibfnamefont {Z.}~\bibnamefont {Li}}, \bibinfo {author} {\bibfnamefont {Q.}~\bibnamefont {Liu}}, \bibinfo {author} {\bibfnamefont {W.}~\bibnamefont {Zhao}}, \bibinfo {author} {\bibfnamefont {K.}~\bibnamefont {Watanabe}}, \bibinfo {author} {\bibfnamefont {T.}~\bibnamefont {Taniguchi}}, \bibinfo {author} {\bibfnamefont {Z.-G.}\ \bibnamefont {Cheng}}, \bibinfo {author} {\bibfnamefont {Z.}~\bibnamefont {Gan}},\ and\ \bibinfo {author} {\bibfnamefont {J.}~\bibnamefont {Lu}},\ }\bibfield  {title} {\bibinfo {title} {Anomalous hall effect in bernal tetralayer graphene enhanced by spin–orbit interaction},\ }\href {https://doi.org/10.1088/1674-1056/adb411} {\bibfield  {journal} {\bibinfo  {journal} {Chinese Physics B}\ }\textbf {\bibinfo
  {volume} {34}},\ \bibinfo {pages} {037201} (\bibinfo {year} {2025}{\natexlab{a}})}\BibitemShut {NoStop}%
\bibitem [{\citenamefont {Qu}\ \emph {et~al.}(2025{\natexlab{b}})\citenamefont {Qu}, \citenamefont {Li}, \citenamefont {Li}, \citenamefont {Hou}, \citenamefont {Han}, \citenamefont {Liu}, \citenamefont {Wang}, \citenamefont {Watanabe}, \citenamefont {Taniguchi}, \citenamefont {Shi},\ and\ \citenamefont {Lu}}]{Qu_2025_2}%
  \BibitemOpen
  \bibfield  {author} {\bibinfo {author} {\bibfnamefont {Z.}~\bibnamefont {Qu}}, \bibinfo {author} {\bibfnamefont {Z.}~\bibnamefont {Li}}, \bibinfo {author} {\bibfnamefont {B.}~\bibnamefont {Li}}, \bibinfo {author} {\bibfnamefont {L.}~\bibnamefont {Hou}}, \bibinfo {author} {\bibfnamefont {X.}~\bibnamefont {Han}}, \bibinfo {author} {\bibfnamefont {Q.}~\bibnamefont {Liu}}, \bibinfo {author} {\bibfnamefont {Z.}~\bibnamefont {Wang}}, \bibinfo {author} {\bibfnamefont {K.}~\bibnamefont {Watanabe}}, \bibinfo {author} {\bibfnamefont {T.}~\bibnamefont {Taniguchi}}, \bibinfo {author} {\bibfnamefont {Y.}~\bibnamefont {Shi}},\ and\ \bibinfo {author} {\bibfnamefont {J.}~\bibnamefont {Lu}},\ }\bibfield  {title} {\bibinfo {title} {Asymmetric gaps of tetralayer graphene unveiled by thermodynamic characterization},\ }\href {https://doi.org/10.1088/1674-1056/adf9fc} {\bibfield  {journal} {\bibinfo  {journal} {Chinese Physics B}\ }\textbf {\bibinfo {volume} {34}},\ \bibinfo {pages} {117201} (\bibinfo {year}
  {2025}{\natexlab{b}})}\BibitemShut {NoStop}%
\bibitem [{\citenamefont {Che}\ \emph {et~al.}(2020)\citenamefont {Che}, \citenamefont {Shi}, \citenamefont {Yang}, \citenamefont {Tian}, \citenamefont {Chen}, \citenamefont {Taniguchi}, \citenamefont {Watanabe}, \citenamefont {Smirnov}, \citenamefont {Lau}, \citenamefont {Shimshoni}, \citenamefont {Murthy},\ and\ \citenamefont {Fertig}}]{PhysRevLett.125.036803}%
  \BibitemOpen
  \bibfield  {author} {\bibinfo {author} {\bibfnamefont {S.}~\bibnamefont {Che}}, \bibinfo {author} {\bibfnamefont {Y.}~\bibnamefont {Shi}}, \bibinfo {author} {\bibfnamefont {J.}~\bibnamefont {Yang}}, \bibinfo {author} {\bibfnamefont {H.}~\bibnamefont {Tian}}, \bibinfo {author} {\bibfnamefont {R.}~\bibnamefont {Chen}}, \bibinfo {author} {\bibfnamefont {T.}~\bibnamefont {Taniguchi}}, \bibinfo {author} {\bibfnamefont {K.}~\bibnamefont {Watanabe}}, \bibinfo {author} {\bibfnamefont {D.}~\bibnamefont {Smirnov}}, \bibinfo {author} {\bibfnamefont {C.~N.}\ \bibnamefont {Lau}}, \bibinfo {author} {\bibfnamefont {E.}~\bibnamefont {Shimshoni}}, \bibinfo {author} {\bibfnamefont {G.}~\bibnamefont {Murthy}},\ and\ \bibinfo {author} {\bibfnamefont {H.~A.}\ \bibnamefont {Fertig}},\ }\bibfield  {title} {\bibinfo {title} {Helical edge states and quantum phase transitions in tetralayer graphene},\ }\href {https://doi.org/10.1103/PhysRevLett.125.036803} {\bibfield  {journal} {\bibinfo  {journal} {Phys. Rev. Lett.}\ }\textbf
  {\bibinfo {volume} {125}},\ \bibinfo {pages} {036803} (\bibinfo {year} {2020})}\BibitemShut {NoStop}%
\bibitem [{\citenamefont {Klanurak}\ \emph {et~al.}(2022{\natexlab{a}})\citenamefont {Klanurak}, \citenamefont {Watanabe}, \citenamefont {Taniguchi}, \citenamefont {Chatraphorn},\ and\ \citenamefont {Taychatanapat}}]{Klanurak20224lg}%
  \BibitemOpen
  \bibfield  {author} {\bibinfo {author} {\bibfnamefont {I.}~\bibnamefont {Klanurak}}, \bibinfo {author} {\bibfnamefont {K.}~\bibnamefont {Watanabe}}, \bibinfo {author} {\bibfnamefont {T.}~\bibnamefont {Taniguchi}}, \bibinfo {author} {\bibfnamefont {S.}~\bibnamefont {Chatraphorn}},\ and\ \bibinfo {author} {\bibfnamefont {T.}~\bibnamefont {Taychatanapat}},\ }\bibfield  {title} {\bibinfo {title} {Magnetoconductance oscillations in electron-hole hybridization gaps and valley splittings in tetralayer graphene},\ }\href {https://doi.org/10.1103/PhysRevB.106.L161405} {\bibfield  {journal} {\bibinfo  {journal} {Phys. Rev. B}\ }\textbf {\bibinfo {volume} {106}},\ \bibinfo {pages} {L161405} (\bibinfo {year} {2022}{\natexlab{a}})}\BibitemShut {NoStop}%
\bibitem [{\citenamefont {Onsager}(1952)}]{Onsager01091952}%
  \BibitemOpen
  \bibfield  {author} {\bibinfo {author} {\bibfnamefont {L.}~\bibnamefont {Onsager}},\ }\bibfield  {title} {\bibinfo {title} {Interpretation of the de haas-van alphen effect},\ }\href {https://doi.org/10.1080/14786440908521019} {\bibfield  {journal} {\bibinfo  {journal} {The London, Edinburgh, and Dublin Philosophical Magazine and Journal of Science}\ }\textbf {\bibinfo {volume} {43}},\ \bibinfo {pages} {1006} (\bibinfo {year} {1952})}\BibitemShut {NoStop}%
\bibitem [{\citenamefont {Lifshitz}\ and\ \citenamefont {Kosevich}(1956)}]{lifshitz1956theory}%
  \BibitemOpen
  \bibfield  {author} {\bibinfo {author} {\bibfnamefont {I.}~\bibnamefont {Lifshitz}}\ and\ \bibinfo {author} {\bibfnamefont {A.~M.}\ \bibnamefont {Kosevich}},\ }\bibfield  {title} {\bibinfo {title} {Theory of magnetic susceptibility in metals at low temperatures},\ }\href {https://doi.org/https://jetp.ras.ru/cgi-bin/dn/e_002_04_0636.pdf} {\bibfield  {journal} {\bibinfo  {journal} {Sov. Phys. JETP}\ }\textbf {\bibinfo {volume} {2}},\ \bibinfo {pages} {636} (\bibinfo {year} {1956})}\BibitemShut {NoStop}%
\bibitem [{\citenamefont {Sahani}\ \emph {et~al.}(2026)\citenamefont {Sahani}, \citenamefont {Das}, \citenamefont {Watanabe}, \citenamefont {Taniguchi}, \citenamefont {Agarwal},\ and\ \citenamefont {Bid}}]{sahani2026quantum}%
  \BibitemOpen
  \bibfield  {author} {\bibinfo {author} {\bibfnamefont {D.}~\bibnamefont {Sahani}}, \bibinfo {author} {\bibfnamefont {S.}~\bibnamefont {Das}}, \bibinfo {author} {\bibfnamefont {K.}~\bibnamefont {Watanabe}}, \bibinfo {author} {\bibfnamefont {T.}~\bibnamefont {Taniguchi}}, \bibinfo {author} {\bibfnamefont {A.}~\bibnamefont {Agarwal}},\ and\ \bibinfo {author} {\bibfnamefont {A.}~\bibnamefont {Bid}},\ }\bibfield  {title} {\bibinfo {title} {Quantum transport spectroscopy of pseudomagnetic field in graphene},\ }\href {https://doi.org/10.1103/vcry-z8kl} {\bibfield  {journal} {\bibinfo  {journal} {Phys. Rev. Lett.}\ }\textbf {\bibinfo {volume} {136}},\ \bibinfo {pages} {166604} (\bibinfo {year} {2026})}\BibitemShut {NoStop}%
\bibitem [{\citenamefont {Chen}\ \emph {et~al.}(2023)\citenamefont {Chen}, \citenamefont {Arora}, \citenamefont {Song},\ and\ \citenamefont {Loh}}]{chen2023gate}%
  \BibitemOpen
  \bibfield  {author} {\bibinfo {author} {\bibfnamefont {H.}~\bibnamefont {Chen}}, \bibinfo {author} {\bibfnamefont {A.}~\bibnamefont {Arora}}, \bibinfo {author} {\bibfnamefont {J.~C.}\ \bibnamefont {Song}},\ and\ \bibinfo {author} {\bibfnamefont {K.~P.}\ \bibnamefont {Loh}},\ }\bibfield  {title} {\bibinfo {title} {Gate-tunable anomalous hall effect in bernal tetralayer graphene},\ }\href {https://doi.org/https://doi.org/10.1038/s41467-023-43796-w} {\bibfield  {journal} {\bibinfo  {journal} {Nature Communications}\ }\textbf {\bibinfo {volume} {14}},\ \bibinfo {pages} {7925} (\bibinfo {year} {2023})}\BibitemShut {NoStop}%
\bibitem [{\citenamefont {Koshino}\ and\ \citenamefont {McCann}(2009)}]{PhysRevB.80.165409}%
  \BibitemOpen
  \bibfield  {author} {\bibinfo {author} {\bibfnamefont {M.}~\bibnamefont {Koshino}}\ and\ \bibinfo {author} {\bibfnamefont {E.}~\bibnamefont {McCann}},\ }\bibfield  {title} {\bibinfo {title} {Trigonal warping and berry's phase $n\ensuremath{\pi}$ in abc-stacked multilayer graphene},\ }\href {https://doi.org/10.1103/PhysRevB.80.165409} {\bibfield  {journal} {\bibinfo  {journal} {Phys. Rev. B}\ }\textbf {\bibinfo {volume} {80}},\ \bibinfo {pages} {165409} (\bibinfo {year} {2009})}\BibitemShut {NoStop}%
\bibitem [{\citenamefont {Avetisyan}\ \emph {et~al.}(2009{\natexlab{b}})\citenamefont {Avetisyan}, \citenamefont {Partoens},\ and\ \citenamefont {Peeters}}]{avetisyan2009electric}%
  \BibitemOpen
  \bibfield  {author} {\bibinfo {author} {\bibfnamefont {A.~A.}\ \bibnamefont {Avetisyan}}, \bibinfo {author} {\bibfnamefont {B.}~\bibnamefont {Partoens}},\ and\ \bibinfo {author} {\bibfnamefont {F.~M.}\ \bibnamefont {Peeters}},\ }\bibfield  {title} {\bibinfo {title} {Electric-field control of the band gap and fermi energy in graphene multilayers by top and back gates},\ }\href {https://doi.org/10.1103/PhysRevB.80.195401} {\bibfield  {journal} {\bibinfo  {journal} {Phys. Rev. B}\ }\textbf {\bibinfo {volume} {80}},\ \bibinfo {pages} {195401} (\bibinfo {year} {2009}{\natexlab{b}})}\BibitemShut {NoStop}%
\bibitem [{\citenamefont {Klanurak}\ \emph {et~al.}(2022{\natexlab{b}})\citenamefont {Klanurak}, \citenamefont {Watanabe}, \citenamefont {Taniguchi}, \citenamefont {Chatraphorn},\ and\ \citenamefont {Taychatanapat}}]{PhysRevB.106.L161405}%
  \BibitemOpen
  \bibfield  {author} {\bibinfo {author} {\bibfnamefont {I.}~\bibnamefont {Klanurak}}, \bibinfo {author} {\bibfnamefont {K.}~\bibnamefont {Watanabe}}, \bibinfo {author} {\bibfnamefont {T.}~\bibnamefont {Taniguchi}}, \bibinfo {author} {\bibfnamefont {S.}~\bibnamefont {Chatraphorn}},\ and\ \bibinfo {author} {\bibfnamefont {T.}~\bibnamefont {Taychatanapat}},\ }\bibfield  {title} {\bibinfo {title} {Magnetoconductance oscillations in electron-hole hybridization gaps and valley splittings in tetralayer graphene},\ }\href {https://doi.org/10.1103/PhysRevB.106.L161405} {\bibfield  {journal} {\bibinfo  {journal} {Phys. Rev. B}\ }\textbf {\bibinfo {volume} {106}},\ \bibinfo {pages} {L161405} (\bibinfo {year} {2022}{\natexlab{b}})}\BibitemShut {NoStop}%
\bibitem [{\citenamefont {Ren}\ \emph {et~al.}(2025)\citenamefont {Ren}, \citenamefont {Shen}, \citenamefont {Xu}, \citenamefont {Shan},\ and\ \citenamefont {Luo}}]{ren2025electric}%
  \BibitemOpen
  \bibfield  {author} {\bibinfo {author} {\bibfnamefont {Y.}~\bibnamefont {Ren}}, \bibinfo {author} {\bibfnamefont {Y.}~\bibnamefont {Shen}}, \bibinfo {author} {\bibfnamefont {C.}~\bibnamefont {Xu}}, \bibinfo {author} {\bibfnamefont {W.}~\bibnamefont {Shan}},\ and\ \bibinfo {author} {\bibfnamefont {W.}~\bibnamefont {Luo}},\ }\bibfield  {title} {\bibinfo {title} {Electric field and doping control of the lifshitz transition in mixed-stacked tetralayer graphene},\ }\href {https://doi.org/10.1103/qbtr-h2sb} {\bibfield  {journal} {\bibinfo  {journal} {Phys. Rev. B}\ }\textbf {\bibinfo {volume} {112}},\ \bibinfo {pages} {115443} (\bibinfo {year} {2025})}\BibitemShut {NoStop}%
\bibitem [{SM()}]{SM}%
  \BibitemOpen
  \href@noop {} {}\bibinfo {note} {Supplementary Material discusses the following: i) Raman characterization; ii) device fabrication and measurement; iii) hole-side Fermi-surface topology; iv) complete zero-field resistivity maps; v) SdH frequency extraction; vi) representative electron-side SdH oscillations; vii) representative hole-side SdH oscillations; viii) SdH frequencies and degeneracies at additional displacement fields; ix) full density--displacement-field response at 1~T; x) high-field transport; and xi) transport responses at high magnetic field.}\BibitemShut {Stop}%
\bibitem [{\citenamefont {Novoselov}\ \emph {et~al.}(2005)\citenamefont {Novoselov}, \citenamefont {Geim}, \citenamefont {Morozov}, \citenamefont {Jiang}, \citenamefont {Katsnelson}, \citenamefont {Grigorieva}, \citenamefont {Dubonos},\ and\ \citenamefont {Firsov}}]{novoselov2005two}%
  \BibitemOpen
  \bibfield  {author} {\bibinfo {author} {\bibfnamefont {K.~S.}\ \bibnamefont {Novoselov}}, \bibinfo {author} {\bibfnamefont {A.~K.}\ \bibnamefont {Geim}}, \bibinfo {author} {\bibfnamefont {S.~V.}\ \bibnamefont {Morozov}}, \bibinfo {author} {\bibfnamefont {D.}~\bibnamefont {Jiang}}, \bibinfo {author} {\bibfnamefont {M.~I.}\ \bibnamefont {Katsnelson}}, \bibinfo {author} {\bibfnamefont {I.~V.}\ \bibnamefont {Grigorieva}}, \bibinfo {author} {\bibfnamefont {S.~V.}\ \bibnamefont {Dubonos}},\ and\ \bibinfo {author} {\bibfnamefont {A.~A.}\ \bibnamefont {Firsov}},\ }\bibfield  {title} {\bibinfo {title} {Two-dimensional gas of massless dirac fermions in graphene},\ }\href {https://doi.org/10.1038/nature04233} {\bibfield  {journal} {\bibinfo  {journal} {nature}\ }\textbf {\bibinfo {volume} {438}},\ \bibinfo {pages} {197} (\bibinfo {year} {2005})}\BibitemShut {NoStop}%
\bibitem [{\citenamefont {Zhang}\ \emph {et~al.}(2005)\citenamefont {Zhang}, \citenamefont {Tan}, \citenamefont {Stormer},\ and\ \citenamefont {Kim}}]{Zhang2005}%
  \BibitemOpen
  \bibfield  {author} {\bibinfo {author} {\bibfnamefont {Y.}~\bibnamefont {Zhang}}, \bibinfo {author} {\bibfnamefont {Y.-W.}\ \bibnamefont {Tan}}, \bibinfo {author} {\bibfnamefont {H.~L.}\ \bibnamefont {Stormer}},\ and\ \bibinfo {author} {\bibfnamefont {P.}~\bibnamefont {Kim}},\ }\bibfield  {title} {\bibinfo {title} {Experimental observation of the quantum hall effect and berry's phase in graphene},\ }\href {https://doi.org/10.1038/nature04235} {\bibfield  {journal} {\bibinfo  {journal} {Nature}\ }\textbf {\bibinfo {volume} {438}},\ \bibinfo {pages} {201} (\bibinfo {year} {2005})}\BibitemShut {NoStop}%
\bibitem [{\citenamefont {Tiwari}\ \emph {et~al.}(2022)\citenamefont {Tiwari}, \citenamefont {Jat}, \citenamefont {Udupa}, \citenamefont {Narang}, \citenamefont {Watanabe}, \citenamefont {Taniguchi}, \citenamefont {Sen},\ and\ \citenamefont {Bid}}]{tiwari2022experimental}%
  \BibitemOpen
  \bibfield  {author} {\bibinfo {author} {\bibfnamefont {P.}~\bibnamefont {Tiwari}}, \bibinfo {author} {\bibfnamefont {M.~K.}\ \bibnamefont {Jat}}, \bibinfo {author} {\bibfnamefont {A.}~\bibnamefont {Udupa}}, \bibinfo {author} {\bibfnamefont {D.~S.}\ \bibnamefont {Narang}}, \bibinfo {author} {\bibfnamefont {K.}~\bibnamefont {Watanabe}}, \bibinfo {author} {\bibfnamefont {T.}~\bibnamefont {Taniguchi}}, \bibinfo {author} {\bibfnamefont {D.}~\bibnamefont {Sen}},\ and\ \bibinfo {author} {\bibfnamefont {A.}~\bibnamefont {Bid}},\ }\bibfield  {title} {\bibinfo {title} {Experimental observation of spin- split energy dispersion in high-mobility single-layer graphene/wse2 heterostructures},\ }\href {https://doi.org/10.1038/s41699-022-00348-y} {\bibfield  {journal} {\bibinfo  {journal} {npj 2D Materials and Applications}\ }\textbf {\bibinfo {volume} {6}},\ \bibinfo {pages} {68} (\bibinfo {year} {2022})}\BibitemShut {NoStop}%
\bibitem [{\citenamefont {Mikitik}\ and\ \citenamefont {Sharlai}(1999)}]{PhysRevLett.82.2147}%
  \BibitemOpen
  \bibfield  {author} {\bibinfo {author} {\bibfnamefont {G.~P.}\ \bibnamefont {Mikitik}}\ and\ \bibinfo {author} {\bibfnamefont {Y.~V.}\ \bibnamefont {Sharlai}},\ }\bibfield  {title} {\bibinfo {title} {Manifestation of berry's phase in metal physics},\ }\href {https://doi.org/10.1103/PhysRevLett.82.2147} {\bibfield  {journal} {\bibinfo  {journal} {Phys. Rev. Lett.}\ }\textbf {\bibinfo {volume} {82}},\ \bibinfo {pages} {2147} (\bibinfo {year} {1999})}\BibitemShut {NoStop}%
\bibitem [{\citenamefont {Jo}\ \emph {et~al.}(2017)\citenamefont {Jo}, \citenamefont {Liu}, \citenamefont {Pfeiffer}, \citenamefont {West}, \citenamefont {Baldwin}, \citenamefont {Shayegan},\ and\ \citenamefont {Winkler}}]{PhysRevB.95.035103}%
  \BibitemOpen
  \bibfield  {author} {\bibinfo {author} {\bibfnamefont {I.}~\bibnamefont {Jo}}, \bibinfo {author} {\bibfnamefont {Y.}~\bibnamefont {Liu}}, \bibinfo {author} {\bibfnamefont {L.~N.}\ \bibnamefont {Pfeiffer}}, \bibinfo {author} {\bibfnamefont {K.~W.}\ \bibnamefont {West}}, \bibinfo {author} {\bibfnamefont {K.~W.}\ \bibnamefont {Baldwin}}, \bibinfo {author} {\bibfnamefont {M.}~\bibnamefont {Shayegan}},\ and\ \bibinfo {author} {\bibfnamefont {R.}~\bibnamefont {Winkler}},\ }\bibfield  {title} {\bibinfo {title} {Signatures of an annular fermi sea},\ }\href {https://doi.org/10.1103/PhysRevB.95.035103} {\bibfield  {journal} {\bibinfo  {journal} {Phys. Rev. B}\ }\textbf {\bibinfo {volume} {95}},\ \bibinfo {pages} {035103} (\bibinfo {year} {2017})}\BibitemShut {NoStop}%
\bibitem [{\citenamefont {Koshino}\ and\ \citenamefont {McCann}(2010)}]{PhysRevB.81.115315}%
  \BibitemOpen
  \bibfield  {author} {\bibinfo {author} {\bibfnamefont {M.}~\bibnamefont {Koshino}}\ and\ \bibinfo {author} {\bibfnamefont {E.}~\bibnamefont {McCann}},\ }\bibfield  {title} {\bibinfo {title} {Parity and valley degeneracy in multilayer graphene},\ }\href {https://doi.org/10.1103/PhysRevB.81.115315} {\bibfield  {journal} {\bibinfo  {journal} {Phys. Rev. B}\ }\textbf {\bibinfo {volume} {81}},\ \bibinfo {pages} {115315} (\bibinfo {year} {2010})}\BibitemShut {NoStop}%
\bibitem [{\citenamefont {Campos}\ \emph {et~al.}(2016)\citenamefont {Campos}, \citenamefont {Taychatanapat}, \citenamefont {Serbyn}, \citenamefont {Surakitbovorn}, \citenamefont {Watanabe}, \citenamefont {Taniguchi}, \citenamefont {Abanin},\ and\ \citenamefont {Jarillo-Herrero}}]{PhysRevLett.117.066601}%
  \BibitemOpen
  \bibfield  {author} {\bibinfo {author} {\bibfnamefont {L.~C.}\ \bibnamefont {Campos}}, \bibinfo {author} {\bibfnamefont {T.}~\bibnamefont {Taychatanapat}}, \bibinfo {author} {\bibfnamefont {M.}~\bibnamefont {Serbyn}}, \bibinfo {author} {\bibfnamefont {K.}~\bibnamefont {Surakitbovorn}}, \bibinfo {author} {\bibfnamefont {K.}~\bibnamefont {Watanabe}}, \bibinfo {author} {\bibfnamefont {T.}~\bibnamefont {Taniguchi}}, \bibinfo {author} {\bibfnamefont {D.~A.}\ \bibnamefont {Abanin}},\ and\ \bibinfo {author} {\bibfnamefont {P.}~\bibnamefont {Jarillo-Herrero}},\ }\bibfield  {title} {\bibinfo {title} {Landau level splittings, phase transitions, and nonuniform charge distribution in trilayer graphene},\ }\href {https://doi.org/10.1103/PhysRevLett.117.066601} {\bibfield  {journal} {\bibinfo  {journal} {Phys. Rev. Lett.}\ }\textbf {\bibinfo {volume} {117}},\ \bibinfo {pages} {066601} (\bibinfo {year} {2016})}\BibitemShut {NoStop}%
\bibitem [{\citenamefont {Slonczewski}\ and\ \citenamefont {Weiss}(1958)}]{Slonczewski1958band}%
  \BibitemOpen
  \bibfield  {author} {\bibinfo {author} {\bibfnamefont {J.~C.}\ \bibnamefont {Slonczewski}}\ and\ \bibinfo {author} {\bibfnamefont {P.~R.}\ \bibnamefont {Weiss}},\ }\bibfield  {title} {\bibinfo {title} {Band structure of graphite},\ }\href {https://doi.org/10.1103/PhysRev.109.272} {\bibfield  {journal} {\bibinfo  {journal} {Phys. Rev.}\ }\textbf {\bibinfo {volume} {109}},\ \bibinfo {pages} {272} (\bibinfo {year} {1958})}\BibitemShut {NoStop}%
\bibitem [{\citenamefont {Koshino}\ and\ \citenamefont {McCann}(2011)}]{PhysRevB.83.165443}%
  \BibitemOpen
  \bibfield  {author} {\bibinfo {author} {\bibfnamefont {M.}~\bibnamefont {Koshino}}\ and\ \bibinfo {author} {\bibfnamefont {E.}~\bibnamefont {McCann}},\ }\bibfield  {title} {\bibinfo {title} {Landau level spectra and the quantum hall effect of multilayer graphene},\ }\href {https://doi.org/10.1103/PhysRevB.83.165443} {\bibfield  {journal} {\bibinfo  {journal} {Phys. Rev. B}\ }\textbf {\bibinfo {volume} {83}},\ \bibinfo {pages} {165443} (\bibinfo {year} {2011})}\BibitemShut {NoStop}%
\bibitem [{\citenamefont {Wu}\ \emph {et~al.}(2015)\citenamefont {Wu}, \citenamefont {Han}, \citenamefont {Lin}, \citenamefont {Zhu}, \citenamefont {He}, \citenamefont {Xu}, \citenamefont {Chen}, \citenamefont {Lu}, \citenamefont {Ye}, \citenamefont {Han}, \citenamefont {Wu}, \citenamefont {Long}, \citenamefont {Shen}, \citenamefont {Huang}, \citenamefont {Wang}, \citenamefont {He}, \citenamefont {Cai}, \citenamefont {Lortz}, \citenamefont {Su},\ and\ \citenamefont {Wang}}]{wu2015detection}%
  \BibitemOpen
  \bibfield  {author} {\bibinfo {author} {\bibfnamefont {Z.}~\bibnamefont {Wu}}, \bibinfo {author} {\bibfnamefont {Y.}~\bibnamefont {Han}}, \bibinfo {author} {\bibfnamefont {J.}~\bibnamefont {Lin}}, \bibinfo {author} {\bibfnamefont {W.}~\bibnamefont {Zhu}}, \bibinfo {author} {\bibfnamefont {M.}~\bibnamefont {He}}, \bibinfo {author} {\bibfnamefont {S.}~\bibnamefont {Xu}}, \bibinfo {author} {\bibfnamefont {X.}~\bibnamefont {Chen}}, \bibinfo {author} {\bibfnamefont {H.}~\bibnamefont {Lu}}, \bibinfo {author} {\bibfnamefont {W.}~\bibnamefont {Ye}}, \bibinfo {author} {\bibfnamefont {T.}~\bibnamefont {Han}}, \bibinfo {author} {\bibfnamefont {Y.}~\bibnamefont {Wu}}, \bibinfo {author} {\bibfnamefont {G.}~\bibnamefont {Long}}, \bibinfo {author} {\bibfnamefont {J.}~\bibnamefont {Shen}}, \bibinfo {author} {\bibfnamefont {R.}~\bibnamefont {Huang}}, \bibinfo {author} {\bibfnamefont {L.}~\bibnamefont {Wang}}, \bibinfo {author} {\bibfnamefont {Y.}~\bibnamefont {He}}, \bibinfo {author} {\bibfnamefont {Y.}~\bibnamefont {Cai}},
  \bibinfo {author} {\bibfnamefont {R.}~\bibnamefont {Lortz}}, \bibinfo {author} {\bibfnamefont {D.}~\bibnamefont {Su}},\ and\ \bibinfo {author} {\bibfnamefont {N.}~\bibnamefont {Wang}},\ }\bibfield  {title} {\bibinfo {title} {Detection of interlayer interaction in few-layer graphene},\ }\href {https://doi.org/10.1103/PhysRevB.92.075408} {\bibfield  {journal} {\bibinfo  {journal} {Phys. Rev. B}\ }\textbf {\bibinfo {volume} {92}},\ \bibinfo {pages} {075408} (\bibinfo {year} {2015})}\BibitemShut {NoStop}%
\bibitem [{\citenamefont {McEllistrim}\ \emph {et~al.}(2023)\citenamefont {McEllistrim}, \citenamefont {Garcia-Ruiz}, \citenamefont {Goodwin},\ and\ \citenamefont {Fal'ko}}]{andrew2023spectroscopic}%
  \BibitemOpen
  \bibfield  {author} {\bibinfo {author} {\bibfnamefont {A.}~\bibnamefont {McEllistrim}}, \bibinfo {author} {\bibfnamefont {A.}~\bibnamefont {Garcia-Ruiz}}, \bibinfo {author} {\bibfnamefont {Z.~A.~H.}\ \bibnamefont {Goodwin}},\ and\ \bibinfo {author} {\bibfnamefont {V.~I.}\ \bibnamefont {Fal'ko}},\ }\bibfield  {title} {\bibinfo {title} {Spectroscopic signatures of tetralayer graphene polytypes},\ }\href {https://doi.org/10.1103/PhysRevB.107.155147} {\bibfield  {journal} {\bibinfo  {journal} {Phys. Rev. B}\ }\textbf {\bibinfo {volume} {107}},\ \bibinfo {pages} {155147} (\bibinfo {year} {2023})}\BibitemShut {NoStop}%
\bibitem [{\citenamefont {Koshino}\ and\ \citenamefont {McCann}(2013)}]{koshino2013multilayer}%
  \BibitemOpen
  \bibfield  {author} {\bibinfo {author} {\bibfnamefont {M.}~\bibnamefont {Koshino}}\ and\ \bibinfo {author} {\bibfnamefont {E.}~\bibnamefont {McCann}},\ }\bibfield  {title} {\bibinfo {title} {Multilayer graphenes with mixed stacking structure: Interplay of bernal and rhombohedral stacking},\ }\href {https://doi.org/10.1103/PhysRevB.87.045420} {\bibfield  {journal} {\bibinfo  {journal} {Phys. Rev. B}\ }\textbf {\bibinfo {volume} {87}},\ \bibinfo {pages} {045420} (\bibinfo {year} {2013})}\BibitemShut {NoStop}%
\bibitem [{\citenamefont {Atri}\ \emph {et~al.}(2024)\citenamefont {Atri}, \citenamefont {Cao}, \citenamefont {Alon}, \citenamefont {Roy}, \citenamefont {Stern}, \citenamefont {Falko}, \citenamefont {Goldstein}, \citenamefont {Kronik}, \citenamefont {Urbakh}, \citenamefont {Hod},\ and\ \citenamefont {Ben~Shalom}}]{atri2024spontaneous}%
  \BibitemOpen
  \bibfield  {author} {\bibinfo {author} {\bibfnamefont {S.~S.}\ \bibnamefont {Atri}}, \bibinfo {author} {\bibfnamefont {W.}~\bibnamefont {Cao}}, \bibinfo {author} {\bibfnamefont {B.}~\bibnamefont {Alon}}, \bibinfo {author} {\bibfnamefont {N.}~\bibnamefont {Roy}}, \bibinfo {author} {\bibfnamefont {M.~V.}\ \bibnamefont {Stern}}, \bibinfo {author} {\bibfnamefont {V.}~\bibnamefont {Falko}}, \bibinfo {author} {\bibfnamefont {M.}~\bibnamefont {Goldstein}}, \bibinfo {author} {\bibfnamefont {L.}~\bibnamefont {Kronik}}, \bibinfo {author} {\bibfnamefont {M.}~\bibnamefont {Urbakh}}, \bibinfo {author} {\bibfnamefont {O.}~\bibnamefont {Hod}},\ and\ \bibinfo {author} {\bibfnamefont {M.}~\bibnamefont {Ben~Shalom}},\ }\bibfield  {title} {\bibinfo {title} {Spontaneous electric polarization in graphene polytypes},\ }\href {https://doi.org/https://doi.org/10.1002/apxr.202300095} {\bibfield  {journal} {\bibinfo  {journal} {Advanced Physics Research}\ }\textbf {\bibinfo {volume} {3}},\ \bibinfo {pages} {2300095} (\bibinfo {year}
  {2024})}\BibitemShut {NoStop}%
\bibitem [{\citenamefont {Wirth}\ \emph {et~al.}(2022)\citenamefont {Wirth}, \citenamefont {Hauck}, \citenamefont {Rothstein}, \citenamefont {Kyoseva}, \citenamefont {Siebenkotten}, \citenamefont {Conrads}, \citenamefont {Klebl}, \citenamefont {Fischer}, \citenamefont {Beschoten}, \citenamefont {Stampfer}, \citenamefont {Kennes}, \citenamefont {Waldecker},\ and\ \citenamefont {Taubner}}]{doi:10.1021/acsnano.2c06053}%
  \BibitemOpen
  \bibfield  {author} {\bibinfo {author} {\bibfnamefont {K.~G.}\ \bibnamefont {Wirth}}, \bibinfo {author} {\bibfnamefont {J.~B.}\ \bibnamefont {Hauck}}, \bibinfo {author} {\bibfnamefont {A.}~\bibnamefont {Rothstein}}, \bibinfo {author} {\bibfnamefont {H.}~\bibnamefont {Kyoseva}}, \bibinfo {author} {\bibfnamefont {D.}~\bibnamefont {Siebenkotten}}, \bibinfo {author} {\bibfnamefont {L.}~\bibnamefont {Conrads}}, \bibinfo {author} {\bibfnamefont {L.}~\bibnamefont {Klebl}}, \bibinfo {author} {\bibfnamefont {A.}~\bibnamefont {Fischer}}, \bibinfo {author} {\bibfnamefont {B.}~\bibnamefont {Beschoten}}, \bibinfo {author} {\bibfnamefont {C.}~\bibnamefont {Stampfer}}, \bibinfo {author} {\bibfnamefont {D.~M.}\ \bibnamefont {Kennes}}, \bibinfo {author} {\bibfnamefont {L.}~\bibnamefont {Waldecker}},\ and\ \bibinfo {author} {\bibfnamefont {T.}~\bibnamefont {Taubner}},\ }\bibfield  {title} {\bibinfo {title} {Experimental observation of abcb stacked tetralayer graphene},\ }\href {https://doi.org/10.1021/acsnano.2c06053}
  {\bibfield  {journal} {\bibinfo  {journal} {ACS Nano}\ }\textbf {\bibinfo {volume} {16}},\ \bibinfo {pages} {16617} (\bibinfo {year} {2022})}\BibitemShut {NoStop}%
\bibitem [{\citenamefont {Xiao}\ \emph {et~al.}(2010)\citenamefont {Xiao}, \citenamefont {Chang},\ and\ \citenamefont {Niu}}]{RevModPhys.82.1959}%
  \BibitemOpen
  \bibfield  {author} {\bibinfo {author} {\bibfnamefont {D.}~\bibnamefont {Xiao}}, \bibinfo {author} {\bibfnamefont {M.-C.}\ \bibnamefont {Chang}},\ and\ \bibinfo {author} {\bibfnamefont {Q.}~\bibnamefont {Niu}},\ }\bibfield  {title} {\bibinfo {title} {Berry phase effects on electronic properties},\ }\href {https://doi.org/10.1103/RevModPhys.82.1959} {\bibfield  {journal} {\bibinfo  {journal} {Rev. Mod. Phys.}\ }\textbf {\bibinfo {volume} {82}},\ \bibinfo {pages} {1959} (\bibinfo {year} {2010})}\BibitemShut {NoStop}%
\bibitem [{\citenamefont {Burgos~Atencia}\ \emph {et~al.}(2024)\citenamefont {Burgos~Atencia}, \citenamefont {Agarwal},\ and\ \citenamefont {Culcer}}]{BurgosAtencia2024}%
  \BibitemOpen
  \bibfield  {author} {\bibinfo {author} {\bibfnamefont {R.}~\bibnamefont {Burgos~Atencia}}, \bibinfo {author} {\bibfnamefont {A.}~\bibnamefont {Agarwal}},\ and\ \bibinfo {author} {\bibfnamefont {D.}~\bibnamefont {Culcer}},\ }\bibfield  {title} {\bibinfo {title} {Orbital angular momentum of bloch electrons: equilibrium formulation, magneto-electric phenomena, and the orbital hall effect},\ }\href {https://doi.org/10.1080/23746149.2024.2371972} {\bibfield  {journal} {\bibinfo  {journal} {Advances in Physics: X}\ }\textbf {\bibinfo {volume} {9}} (\bibinfo {year} {2024})}\BibitemShut {NoStop}%
\end{thebibliography}%

\section{End Matter}
\subsection{EM1: Tight-binding description of B-4LG} \phantomsection
\label{4lg_tb}

\begin{figure}[h!]
    \centering
    \includegraphics[width=0.85\linewidth]{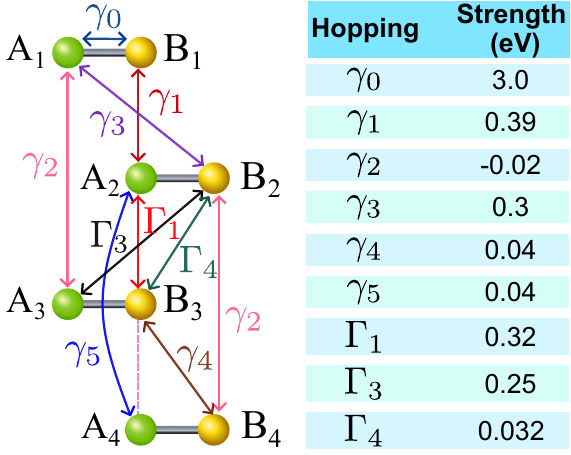}
    \caption{\justifying Schematic representation of the Slonczewski--Weiss--McClure tight-binding model for Bernal-stacked tetralayer graphene. Left: sublattice-resolved lattice structure showing the dominant hopping processes. The parameters $\gamma_0$, $\gamma_1$, $\gamma_3$, and $\gamma_4$ describe the intralayer and nearest-layer couplings within the two bilayer-like subsystems; $\gamma_2$ and $\gamma_5$ describe next-nearest-layer couplings; and $\Gamma_1$, $\Gamma_3$, and $\Gamma_4$ describe the couplings across the central layer interface. Right: numerical values of the hopping parameters used throughout this work.}
    \label{fig_4LG_EM}
\end{figure}

Bernal-stacked tetralayer graphene (B-4LG) consists of four graphene layers arranged in the ABAB stacking sequence, with eight carbon atoms per unit cell, as illustrated in Fig.~\ref{fig_4LG_EM}. From a low-energy perspective, B-4LG may be viewed as two Bernal bilayer graphene (BLG) subsystems coupled via interbilayer hopping. We describe its electronic structure using the Slonczewski--Weiss--McClure tight-binding model~\cite{Slonczewski1958band, PhysRevB.83.165443,wu2015detection,andrew2023spectroscopic,koshino2013multilayer}. The basis is chosen $\Psi=(A_1,B_1,A_2,B_2,A_3,B_3,A_4,B_4)^T$,
where $A_i$ and $B_i$ denote the two sublattices in layer $i$.

The relevant hopping processes and their numerical values are summarized in Fig.~\ref{fig_4LG_EM}. The intralayer nearest-neighbor hopping $\gamma_0$ determines the Dirac velocity of an isolated graphene sheet. Within each BLG subsystem, the dominant interlayer dimer hopping $\gamma_1$ generates the characteristic bilayer-like parabolic dispersion, while the skew hoppings $\gamma_3$ and $\gamma_4$ introduce trigonal warping and electron--hole asymmetry. The next-nearest-layer couplings $\gamma_2$ and $\gamma_5$ further modify the low-energy spectrum and shift the relative energies of the low-energy bands.

The two BLG subsystems are coupled through the inter-bilayer hopping amplitudes $\Gamma_1$, $\Gamma_3$, and $\Gamma_4$. Here, $\Gamma_1$ corresponds to the vertical dimer hopping between the two BLGs, whereas $\Gamma_3$ and $\Gamma_4$ represent skew inter-bilayer hoppings. These couplings hybridize the electronic states of the two BLGs and are ultimately responsible for the coexistence of light- and heavy-mass bilayer-like bands near charge neutrality.

A perpendicular displacement field is incorporated through layer-dependent electrostatic potentials. This field breaks inversion symmetry and lifts the degeneracy of the low-energy bands, leading to strong band hybridization and the opening of energy gaps at the band-touching points. The resulting Hamiltonian can be written in block form as~\cite{atri2024spontaneous, doi:10.1021/acsnano.2c06053, andrew2023spectroscopic, PhysRevLett.120.096802, doi:10.1021/acs.nanolett.4c01133}
\begin{equation}
\begin{aligned}
 {\mathcal H}(\kb) &=
\begin{pmatrix}
{\cal H}^{12}_{\rm BLG}(\kb) & \mathbb{V}^\dagger(\kb) \\
\mathbb{V}(\kb) & {\cal H}^{34}_{\rm BLG}(\kb)
\end{pmatrix},
\end{aligned}
\end{equation}
where ${\cal H}^{12}_{\rm BLG}$ and ${\cal H}^{34}_{\rm BLG}$ describe the top and bottom BLG subsystems, respectively, and $\mathbb{V}$ represents the inter-bilayer coupling matrix. These matrices are defined as
\begin{equation}
{\cal H}_{\rm BLG}^{12}(\kb) =
\begin{pmatrix}
\Delta & v_0\,\pi^\dagger & -v_4\,\pi^\dagger & v_3\, \pi \\

v_0\, \pi & \Delta + \Delta' & \gamma_1 & -v_4\,\pi^\dagger \\

-v_4\, \pi  & \gamma_1 & \frac{\Delta}{3} + \Delta' & v_0\, \,\pi^\dagger \\
v_3\, \pi^\dagger &  -v_4\, \pi & v_0\, \pi & \frac{\Delta}{3}
\end{pmatrix}~,
\end{equation}
\begin{equation}
{\cal H}^{34}_{\rm BLG}(\kb) =
\begin{pmatrix}
-\frac{\Delta}{3} & v_0\,\pi^\dagger & -v_4\,\pi^\dagger & v_3\, \pi \\

v_0\, \pi & -\frac{\Delta}{3} + \Delta' & \gamma_1 & -v_4\,\pi^\dagger \\

-v_4\, \pi  & \gamma_1 & -\Delta + \Delta' & v_0\, \,\pi^\dagger \\
v_3\, \pi^\dagger &  -v_4\, \pi & v_0\, \pi & -\Delta
\end{pmatrix}~,
\end{equation}
and
\begin{equation}
\mathbb{V}(\kb) =
\begin{pmatrix}
\frac{\gamma_2}{2} & 0 & -V_4\, \pi^\dagger &  V_3\, \pi \\
0 & \frac{\gamma_5}{2} & \Gamma_1 & -V_4\, \pi^\dagger \\
0  & 0 & \frac{\gamma_5}{2} & 0  \\
0 &  0 & 0 & \frac{\gamma_2}{2} 
\end{pmatrix}.
\end{equation}
Here, $\pi=\xi p_x + i p_y$, where $\xi=\pm1$ labels the $K$ and $K'$ valleys. The velocity parameters are defined as
\begin{equation}
v_i=\frac{\sqrt{3}a\gamma_i}{2\hbar},
\qquad
V_i=\frac{\sqrt{3}a\Gamma_i}{2\hbar},
\end{equation}
where $a=0.246~{\rm nm}$ is the graphene lattice constant. We use the layer-potential convention $(U_1,U_2,U_3,U_4)=(\Delta,\Delta/3,-\Delta/3,-\Delta)$, so that $2\Delta$ is the onsite-energy difference between the outermost layers. The field-induced onsite-energy scale is written as $\Delta=eDd_{\rm eff}/(\epsilon_0\epsilon_r)$, where $D/\epsilon_0$ is the applied displacement field, $\epsilon_r \approx 5.8$ is the effective dielectric constant~\cite{PhysRevLett.120.096802}, and $d_{\rm eff}$ is half the outer-layer separation; for equal interlayer spacing $d_0$, $d_{\rm eff}=3d_0/2$. In contrast, $\Delta'$ denotes the intrinsic onsite-energy difference between dimer and non-dimer sublattice sites, introducing a weak electron--hole asymmetry into the spectrum. While $\Delta$ is externally tunable, $\Delta'$ is a fixed material parameter determined by the local crystal environment. Throughout this work, we use $\Delta' = 0.0303$~eV.


\subsection{EM2: Valley-orbital origin of the large valley splitting} \label{sec:valley_orbital_mechanism}

\begin{figure*}[t!]
    \centering
    \includegraphics[width=0.9\linewidth]{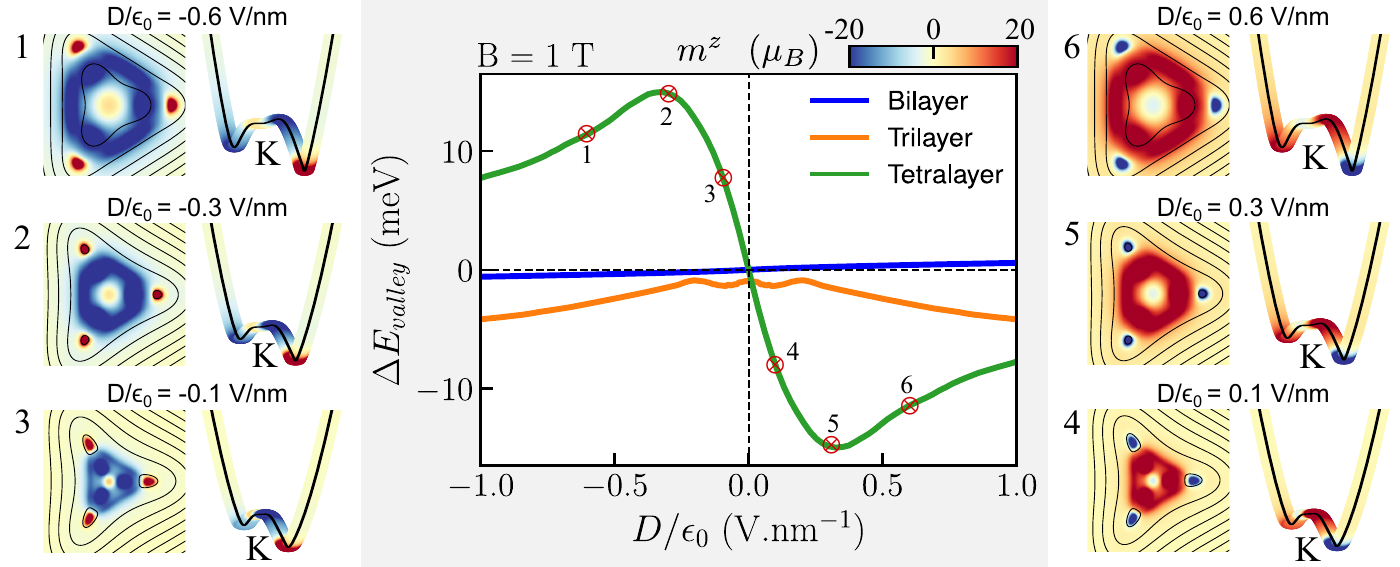}
    \caption{ \justifying
    Orbital-Zeeman-induced valley energy splitting as a function of displacement field for Bernal bilayer graphene (AB BLG), Bernal trilayer graphene (ABA TLG), and Bernal tetralayer graphene (ABAB B-4LG) at $B_\perp=1~\mathrm{T}$. The splitting is estimated from the orbital magnetic moment (OMM) at the relevant conduction-band minimum. B-4LG exhibits a large, sign-changing splitting caused by the enhanced OMM of its hybridized low-energy bands. Insets (1--6) show the momentum-space distribution of the OMM, overlaid with black constant-energy contours, and the corresponding low-energy dispersion weighted by the OMM near the $K$ valley at the marked displacement fields.
}
    \label{fig:omm_splitting}
\end{figure*}

The large valley splitting observed in B-4LG is not a generic consequence of increasing the number of graphene layers. It originates from the specific multiband structure of ABAB tetralayer graphene, in which light- and heavy-mass bilayer-like bands lie close in energy and are strongly coupled through inter-bilayer hopping. At $D=0$, inversion symmetry relates the $K$ and $K'$ valleys and protects their degeneracy in the single-particle spectrum. A finite displacement field breaks inversion symmetry and polarizes the low-energy wave functions across the four layers. A perpendicular magnetic field also breaks time-reversal symmetry, allowing the valley-contrasting orbital magnetic moments to shift the two valleys in opposite directions.

Within the semiclassical wave-packet picture, the orbital magnetic moment of band $n$ is calculated from the tight-binding Bloch states as~\cite{RevModPhys.82.1959, BurgosAtencia2024}
\begin{equation}
{\bm m}_n({\bm k})=\frac{e}{2\hbar}\operatorname{Im}
\left\langle \nabla_{\bm k}u_{n{\bm k}}\right|
\times\left[\mathcal H({\bm k})-\varepsilon_{n{\bm k}}\right]
\left|\nabla_{\bm k}u_{n{\bm k}}\right\rangle .
\end{equation}
where $e>0$ is the elementary charge. The orbital-Zeeman correction to the band energy is $\widetilde{\varepsilon}_{n{\bm k}}=\varepsilon_{n{\bm k}}-{\bm m}_n({\bm k})\!\cdot\!{\bm B}$. Because time-reversal-related valleys carry opposite orbital moments, ${\bm m}_{K'}=-{\bm m}_{K}$, we define the signed splitting plotted in Fig.~\ref{fig:omm_splitting} as
\begin{equation}
\begin{aligned}
\Delta E_{\rm valley}&\equiv E_K-E_{K'}
\simeq-2m_{K,z}(D)B_\perp,\\
\left|\Delta E_{\rm valley}\right|&\simeq
2\left|m_{K,z}(D)\right|B_\perp.
\end{aligned}
\end{equation}
Here, $m_{K,z}(D)$ is the out-of-plane OMM evaluated at the $K$-valley conduction-band minimum that evolves into the lowest electron-like Landau levels.

To quantify this mechanism, we calculate $\Delta E_{\rm valley}$ for AB BLG, ABA TLG, and ABAB B-4LG at the same magnetic field and using the same electrostatic convention [Fig.~\ref{fig:omm_splitting}]. BLG and TLG exhibit weak splitting across the investigated displacement-field range, whereas B-4LG develops a large, sign-reversing splitting that substantially exceeds those of the lower-layer systems. The six marked points show the momentum-space OMM and the corresponding low-energy dispersion weighted by the OMM strength. The enhancement is concentrated near avoided crossings between the light- and heavy-mass bilayer-like bands and near the trigonal-warped band extrema, where strong interband hybridization produces large Berry curvature and orbital magnetic moment. At $B_\perp=1~\mathrm{T}$ and $D/\epsilon_0\sim0.6~\mathrm{V\,nm^{-1}}$, the orbital-Zeeman estimate reaches the several-meV scale, of the same order of magnitude as the $\sim6~\mathrm{meV}$ splitting obtained from the full LL calculation. The two values need not coincide exactly because the semiclassical estimate is evaluated from the zero-field band edge, whereas the LL calculation includes orbital quantization. These results identify orbital-Zeeman coupling as the dominant single-particle origin of the large valley splitting in B-4LG.

\end{document}